\documentclass[aps,prb,twocolumn,floatfix,superscriptaddress,10pt,amsfonts,amssymb,amsmath]{revtex4-2}
\usepackage[utf8]{inputenc}
\usepackage[T1]{fontenc}
\usepackage[english]{babel}
\usepackage{microtype}
\usepackage[varbb,varg,otfmath]{newtx} %

\usepackage{enumitem,tabularx}
\usepackage{graphicx}
\usepackage[dvipsnames,svgnames,x11names]{xcolor}
\usepackage{hyperref}
\hypersetup{ %
    colorlinks=true, linkcolor=NavyBlue, urlcolor=NavyBlue, citecolor=NavyBlue
}
\usepackage{physics,braket}
\usepackage[capitalise]{cleveref}
\crefname{figure}{Fig.}{Figs.}
\Crefname{figure}{Fig.}{Figs.}
\crefname{equation}{Eq.}{Eqs.}
\Crefname{equation}{Eq.}{Eqs.}
\crefname{section}{Sec.}{Secs.}
\Crefname{section}{Sec.}{Secs.}
\crefname{appendix}{App.}{Apps.}
\Crefname{appendix}{App.}{Apps.}
\crefname{table}{Table}{Tables}
\Crefname{table}{Table}{Tables}
\AtBeginDocument{}
\usepackage{bm,dsfont,eucal}
\usepackage{tikz}

\AtBeginDocument{\renewcommand{\vec}[1]{\bm{#1}}}

\begin{document}

\title{Quantum Monte-Carlo study of spin-valley ferromagnetism in rhombohedral graphene}

\author{Michael Demos}
\affiliation{Department of Physics and Astronomy, University of Kentucky, Lexington, Kentucky 40506-0055, USA}
\author{Dongqi Li}
\affiliation{Department of Physics and Astronomy, University of Kentucky, Lexington, Kentucky 40506-0055, USA}
\author{Mainak Das}
\affiliation{Department of Physics and Astronomy, University of Kentucky, Lexington, Kentucky 40506-0055, USA}
\author{Tobias\ M.\ R.\ Wolf}
\affiliation{Department of Physics, The University of Texas at Austin, Austin, Texas 78712, USA}
\author{Chunli Huang}
\affiliation{Department of Physics and Astronomy, University of Kentucky, Lexington, Kentucky 40506-0055, USA}

\date{\today} 
\begin{abstract}
Recent capacitance measurements in bilayer and trilayer rhombohedral graphene reported large magnetic entropy above the Curie temperature, suggesting the formation of fluctuating local moments. Further experimental support comes from scanning tunneling microscopy experiments in rhombohedral hexalayer graphene, which directly visualized honeycomb electron-crystal order that strengthens upon heating before melting, related to the  Pomeranchuk effect.
Motivated by these experiments, we construct spin-valley local-moment Hamiltonians for the quarter-metal and half-metal phases. We extract their couplings from constrained Hartree--Fock calculations, and solve their finite-temperature thermodynamics with stochastic-series-expansion quantum Monte Carlo. For the quarter metal we find a continuous 2D-Ising transition at $T_c \approx 0.7$~K, an order of magnitude smaller than the Hartree--Fock estimate ($\sim 30$~K) and closer to the experimental scale of a few kelvin. 
For the half metal, we study the field-induced magnetization as a function of temperature with and without proximity-enhanced spin--orbit coupling. When spin--orbit coupling is negligible, intervalley Hund's coupling enhances the magnetic response. Strong spin--orbit coupling, as can be induced by a neighboring transition-metal dichalcogenide, gives rise to the standard spin-canted ground state at zero temperature. Upon heating, the in-plane polarization decreases more rapidly than the out-of-plane polarization. Surprisingly, despite the itinerant nature of the electrons, local-moment models with microscopically determined couplings appear to capture key finite-temperature magnetic properties of rhombohedral graphene without any fitting.
\end{abstract}

\maketitle

\section{Introduction}

Since the discovery of itinerant-electron ferromagnetism in the form of half-metal and quarter-metal phases in rhombohedral multilayer graphene \cite{zhou_half_2021}, this family of systems has emerged as a versatile platform for correlated-electron phenomena, ranging from orbital multiferroicity \cite{Han2023} and the fractional quantum anomalous Hall \cite{Lu2024} effect to unconventional chiral superconductivity \cite{Han2025,Geier2026} as well as many other exotic phases \cite{Ghazaryan2023, Slizovskiy2019, Zhou_superconductivity_2021, Han2024, Yin2022}. The simplest microscopic description, Hartree--Fock theory built on the full Slonczewski--Weiss--McClure band structure together with long-range Coulomb interactions, reproduces the experimentally observed density-displacement-field phase diagram and predicts ferromagnetic half- and quarter-metallic states \cite{huang2023spin,zhou_half_2021}.  Quantitative agreement, however, requires an unusually large screening constant (e.g.\ $\varepsilon \approx 20$ instead of the bare $\varepsilon \approx 4$ expected from hexagonal boron nitride), reducing the Coulomb energy so that it is comparable to the Fermi energy.  Despite this ad hoc parameter choice, the approach has been quite successful. Magneto-transport measurements unambiguously associate the onset of ferromagnetism with a reduction of Fermi-surface degeneracy, indicating that exchange energy, accurately described by the mean-field theory, plays the dominant role.

However, recent experiments by Holleis \emph{et al.}~\cite{Holleis_2025} reported finite-temperature behavior that indicates this picture might not be complete.  Above the Curie temperature, where long-range order disappears, the paramagnetic metal retains an unusually large entropy of order $S \sim k_{\mathrm B}$ per carrier, persisting to temperatures nearly an order of magnitude higher than $T_c$.  The data are consistent with a good metal hosting fluctuating local moments, an analogue of the almost-ferromagnetic Fermi liquid observed near the melting curve of $^3$He~\cite{doniach1966low,vollhardt1984normal,georges1997normal}.  Importantly, the system remains a low-resistance conductor even while carrying such a large magnetic entropy, very different from conventional local-moment materials in which the charge sector is insulating.

\begin{figure}[!t]
\centering
\includegraphics[width=1\linewidth]{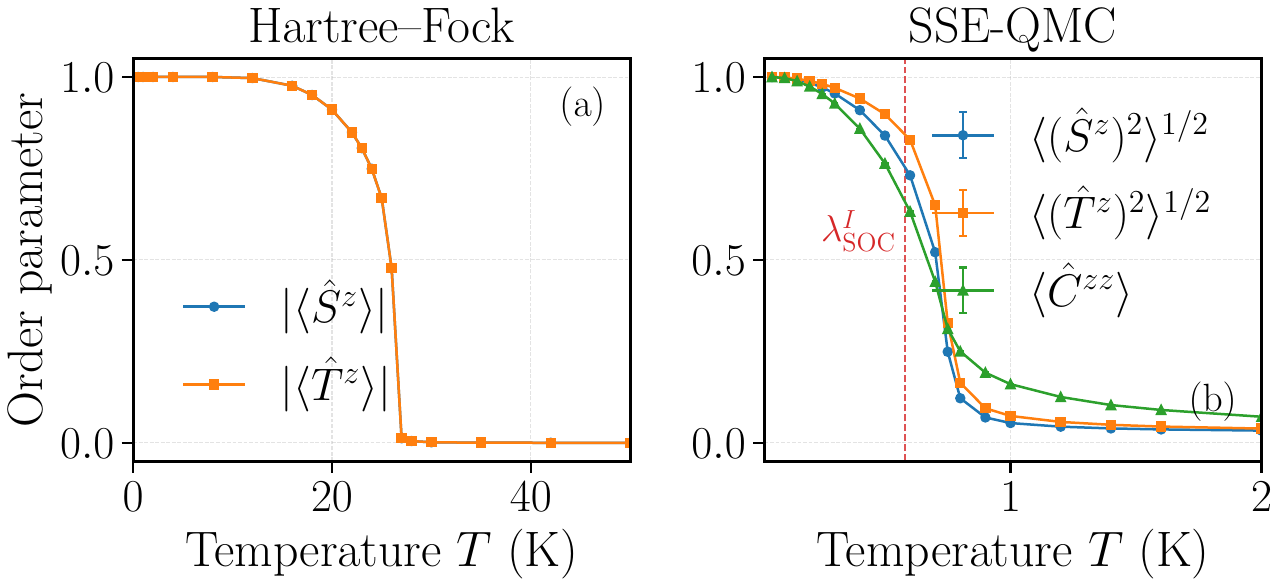}
\caption{Finite-temperature quarter-metal (QM) order.
(a) Self-consistent Hartree--Fock for the full continuum model gives $T_c^{\rm HF}\sim 30$~K. 
(b) Stochastic-series-expansion quantum Monte Carlo for the projected local-moment Hamiltonian at $L=40$ gives a transition near $T_c\approx0.7$~K in the RMS spin polarization $\langle(\hat S^z)^2\rangle^{1/2}$, RMS valley polarization $\langle(\hat T^z)^2\rangle^{1/2}$, and local spin-valley locking $\langle\hat C^{zz}\rangle$. Polarizations are dimensionless in the saturation-1 convention of \cref{sec:local_moment}. The locking observable is invariant under the broken joint $\mathbb Z_2$ flip and is therefore insensitive to large uncertainties arising from domain fluctuations in the simulation. The dashed line marks graphene's intrinsic spin--orbit scale.
}
\label{fig:MF_vs_QMC}
\end{figure}

A recent scanning tunneling microscopy (STM) experiment~\cite{sun2026real} has directly visualized long-range charge order in rhombohedral hexalayer graphene. Interestingly, the honeycomb charge-order pattern becomes more pronounced in the $dI/dV$ maps upon warming from 0.3~K to 1.6~K. The authors suggest that this behavior is reminiscent of the Pomeranchuk effect, in which entropy associated with internal degrees of freedom favors electron crystallization at higher temperatures.

Mean-field theory fails to describe these finite-temperature properties (\cref{fig:MF_vs_QMC}): self-consistent Hartree--Fock places the quarter-metal Curie temperature at $T_c^{\rm MF}\sim 30$~K, more than an order of magnitude above the few-kelvin scale of experiment~\cite{Holleis_2025}. The reason is well known, but worth emphasizing. Hartree--Fock theory describes thermal suppression of the order parameter primarily through a reduction of its amplitude: majority spin-valley flavors are depopulated while minority spin-valley flavors are populated. The relevant low-energy excitations, however, are not amplitude fluctuations, but transverse rotations of the order parameter. These transverse collective modes can be captured at zero temperature within the time-dependent Hartree--Fock approximation, especially in the long-wavelength limit~\cite{wolf2408magnetism}. At elevated temperatures, however, even time-dependent Hartree--Fock theory breaks down. This limitation is familiar from the theory of metallic ferromagnetism and has been discussed extensively in reviews such as Ref.~\cite{moriya1979recent}.

A spin-canted state arising from the competition between proximity-enhanced Ising spin--orbit coupling and intervalley exchange has recently been discussed in rhombohedral graphene with transition-metal dichalcogenide substrate~\cite{patterson2025superconductivity}. Theoretical work has further proposed that low-energy fluctuations of the spin-canted state can mediate superconducting pairing~\cite{dong2026superconductivity}. These results also motivate us to examine how the canted state evolves with temperature from local moment models.

Motivated by these works, we construct effective \emph{local-moment} Hamiltonians that retain the spin-valley degrees of freedom with all couplings extracted from self-consistent Hartree--Fock calculations. 
We then use the stochastic-series-expansion quantum Monte Carlo method to obtain numerically exact thermodynamic properties. Our main results are as follows. (1) For the quarter-metal phase, we find a second-order Ising transition at $T_c \approx 0.7$~K, more than an order of magnitude smaller than the full Hartree--Fock estimate (${\sim30}$~K) and of the same order as experiment~\cite{Holleis_2025}, $T_c^{\rm experiment} \approx 3.0~{\rm K}$. The Hartree--Fock transition is first order while the quantum Monte Carlo transition is continuous. (2) For the half-metal phase with TMD-proximity-enhanced spin--orbit coupling, we find a finite-temperature Berezinskii--Kosterlitz--Thouless transition near $T_{\rm KT}\approx 2.0~{\rm K}$ in a valley-canted antiferromagnetic regime.

These results show that local-moment models provide a useful starting point to understand the finite-temperature magnetism of rhombohedral trilayer graphene. The remainder of the paper is organized as follows: \cref{sec:local_moment} introduces the local-moment Hamiltonians; \cref{sec:QMC} outlines the stochastic-series-expansion quantum Monte Carlo method; \cref{sec:results} presents the numerical results; \cref{sec:summary} compares with experiment.

\begin{figure*}
\centering
\includegraphics[width=\linewidth]{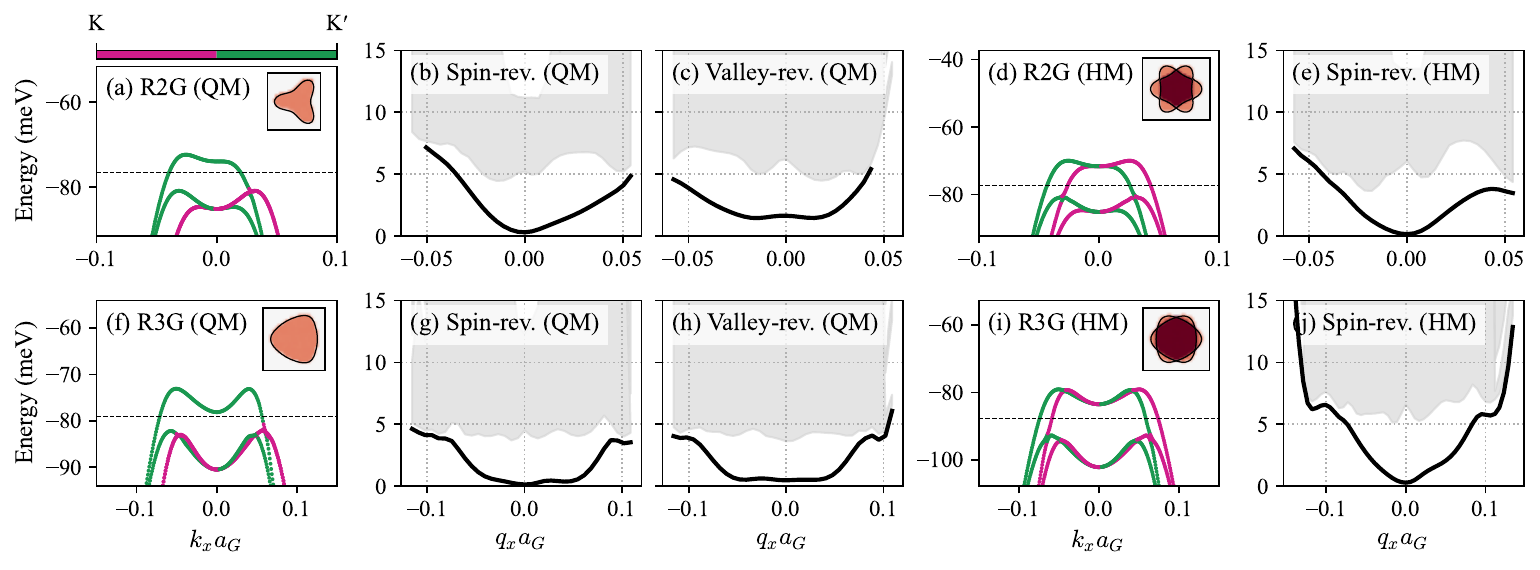}
\caption{%
Single-particle and collective-mode spectra of rhombohedral bilayer (R2G, a–e) and trilayer graphene (R3G, f–j). $a_G=0.246$~nm is the graphene lattice constant. 
(a,d,f,i) Hartree--Fock band structures of half-metal (HM) and quarter-metal (QM) flavor-polarized states, along $k_x$ and colored by valley ($K$, $K'$).  The dashed line marks the chemical potential and insets show the Fermi surface.  
(b,c,e,g,h,j) Flavor-reversal excitation spectra for each corresponding HF state in time-dependent Hartree--Fock approximation, along $q_x$, see Ref.~\cite{wolf2408magnetism} for details. The collective low-energy modes are shown as solid lines and the particle-hole continuum as a gray shaded region; for the HM case the spin-reversal spectra (e,j) are valley-projected.
Top row (a–e) uses carrier densities $n_e^{\scriptscriptstyle\text{QM}}=-1.1\times10^{11}\,\mathrm{cm^{-2}}$ and $n_e^{\scriptscriptstyle\text{HM}}=-3.0\times10^{11}\,\mathrm{cm^{-2}}$.
Bottom row (f–j) uses $n_e^{\scriptscriptstyle\text{QM}}=-5.2\times10^{11}\,\mathrm{cm^{-2}}$ and $n_e^{\scriptscriptstyle\text{HM}}=-1.12\times10^{12}\,\mathrm{cm^{-2}}$.
We used interlayer potential $U_D=60\,$meV, dual-gate static screening ($d_{\text{gate}}=40\,$nm) and effective dielectric constants $\varepsilon_r^{\scriptscriptstyle\text{R2G}}=13$, $\varepsilon_r^{\scriptscriptstyle\text{R3G},\text{HM}}=18$, $\varepsilon_r^{\scriptscriptstyle\text{R3G},\text{QM}}=25$. The slightly smaller $\varepsilon_r=15$ used in the parameter-extraction calculations of \cref{appendix:parameter_estimation} reflects a different gating geometry ($d_{\text{gate}}=50\,$nm) and does not qualitatively affect the extracted couplings. 
}
\label{fig:R2G_R3G_TDHF}
\end{figure*}

\section{Local-Moment Models}
\label{sec:local_moment}

Before introducing local-moment models, we first summarize what is known from Hartree--Fock and time-dependent Hartree--Fock calculations. Within the Hartree--Fock approximation, the quarter-metal phase occupies one of the four spin-valley flavors, while the half-metal phase occupies two of the four. The magnitude of flavor polarization is well captured at the mean-field level, although the orientation of the polarization in spin-valley space depends on short-range interactions and spin--orbit coupling. In this paper, we focus on a spin- and valley-polarized quarter metal rather than an intervalley-coherent state, and on a spin-polarized but valley-unpolarized half metal. These states are the dominant phases in the phase diagrams of rhombohedral bilayer (R2G) and trilayer (R3G) graphene. The corresponding band structures are shown in \Cref{fig:R2G_R3G_TDHF}(a,d,f,i).

Time-dependent Hartree--Fock calculations at zero temperature show three distinct collective modes: valley-reversal, spin-reversal, and combined spin-valley-reversal modes. A common feature of these modes is that they are spectrally separated from the Stoner continuum over a broad range of wavevectors, extending up to momenta comparable to the Fermi momentum~\cite{wolf2408magnetism}. The corresponding results for bilayer and trilayer graphene are shown in \Cref{fig:R2G_R3G_TDHF}. This low-energy spectral separation over a momentum range comparable to the Fermi wavevector is an important observation. It suggests that a local-moment description may be appropriate. The flavor-rotation degrees of freedom form well-defined low-energy bands over a broad range of wavevectors while remaining largely decoupled from charge fluctuations in the Stoner continuum.

This structure is reminiscent of the well-known $\nu=1$ quantum Hall ferromagnet. Within the lowest-Landau-level projection, the single spin-reversal spectrum obtained from time-dependent Hartree--Fock is exact, because at each momentum $q$ there is only one propagating spin-flip state. The $q=0$ mode corresponds to Zeeman-gapped Larmor precession, and the excitation energy increases with $q$, approaching the Hartree--Fock exchange gap at large momenta. Nevertheless, even in this comparatively simple case, exact finite-temperature results are not available analytically. Its finite-temperature properties are instead well described by stochastic-series-expansion quantum Monte Carlo methods~\cite{henelius2000monte,henelius2002quantum}.

We follow a similar logic to describe finite-temperature properties of a spin- and valley-polarized quarter metal. Specifically, we introduce a model describing the relevant long-wavelength fluctuations of the spin-valley-polarized ground state,
\begin{equation}
\label{eq:H_QM_main}
\hat H_{1/4}=-J_s \!\! \sum_{\langle i,j \rangle} \! \hat{\vec{s}}_i \cdot \hat{\vec{s}}_j
-J_\tau \!\! \sum_{\langle i,j \rangle} \! \hat{\vec{\tau}}_i \cdot \hat{\vec{\tau}}_j
-\frac{\lambda}{2} \! \sum_i \hat s^z_i \hat\tau^z_i .
\end{equation}
Here $\hat{\vec{s}}_i$ and $\hat{\vec{\tau}}_i$ are Pauli-vector operators acting in the spin and valley spaces at site $i$, respectively, normalized such that a fully polarized state has $|\langle\hat{\vec{s}}_i\rangle|=|\langle\hat{\vec{\tau}}_i\rangle|=1$. Throughout, operators carry hats; unhatted symbols always denote numbers (eigenvalues, sampled configuration values, couplings, or classical order-parameter fields). The sum $\langle i,j\rangle$ runs over nearest-neighbor bonds of the square lattice, with each bond counted once. The parameters $J_s$ and $J_\tau$ are the spin and valley stiffnesses extracted from Hartree--Fock calculations, with $J_s,J_\tau>0$ favoring ferromagnetic spin and valley alignment. This is consistent with the general principle that exchange between orthogonal Bloch states in the presence of repulsive Coulomb interactions favors ferromagnetism~\cite{herring1966exchange}. We place the moments on an $L\times L$ square lattice with $N=L^2$ sites. The precise lattice geometry enters only through the mapping between the continuum stiffness and the lattice exchange couplings (\cref{appendix:parameter_estimation}); the universal long-wavelength physics is insensitive to these microscopic details.
For $\lambda=0$, $\hat H_{1/4}$ is invariant under independent spin and valley rotations, $SU(2)_s\times SU(2)_\tau$. Non-zero spin--orbit coupling reduces the symmetry to $U(1)_s\times U(1)_\tau \times Z_2$. This gives rise to a finite temperature phase transition to a spin and valley polarized state in the presence of zero external field, which is absent in the conventional 2D Heisenberg model.

For the half-metal phase, experimental data~\cite{zhou_half_2021} indicate full spin polarization but no valley polarization. We therefore propose an effective local-moment model in which spin fluctuations occur in both valleys at fixed valley occupation. The corresponding Hamiltonian is
\begin{align}
\label{eq:H_HM_main}
\hat H_{1/2} &=
-J_{s}\!\!\sum_{\langle i,j\rangle}\!
\bigl(\hat{\vec s}_{i,K}\!\cdot\!\hat{\vec s}_{j,K}
+ \hat{\vec s}_{i,\bar K}\!\cdot\!\hat{\vec s}_{j,\bar K}\bigr)
-g_{\perp}\sum_{i} \hat{\vec s}_{i,K}\!\cdot\!\hat{\vec s}_{i,\bar K} \nonumber\\
&\quad
-\frac{\lambda}{2}\sum_{i}
\bigl(\hat s_{i,K}^z-\hat s_{i,\bar K}^z\bigr)
-h_{x}\sum_{i}
\bigl(\hat s_{i,K}^x+\hat s_{i,\bar K}^x\bigr).
\end{align}
Here $\hat{\vec{s}}_{i,K}$ and $\hat{\vec{s}}_{i,\bar K}$ are Pauli-vector spin operators in valleys $K$ and $\bar K$, respectively, with the same normalization as above. The intervalley coupling $g_\perp>0$ favors ferromagnetic alignment across valleys. The strong displacement field in experiment~\cite{wolf2408magnetism} is expected to suppress an attractive electron-phonon contribution to $g_\perp$. Although intervalley-coherent ordering tendencies have been reported in rhombohedral graphene, we restrict our attention here to situations where valley is a good quantum number and focus on spin fluctuations within the valley.

At $h_x=0$, the competition between intervalley exchange and valley-dependent Ising spin--orbit coupling produces a canted ground state for $0<\lambda<4g_\perp$, characterized by a common in-plane spin component and opposite out-of-plane components in the two valleys. Classical minimization gives $\cos\theta=\lambda/(4g_\perp)$ (\cref{appendix:hm_g_dependence}).
\cref{eq:H_HM_main} has a hidden combined valley and spin symmetry. Let $\hat{\mathcal P}=\hat P_{K\bar K}\hat U_x(\pi)$, where $\hat P_{K\bar K}$ exchanges the two valleys and $\hat U_x(\pi)$ rotates all spins by $\pi$ around the $x$ axis. This operation maps
\begin{equation}
\label{eq:HM_hidden_symmetry}
    \hat s_{i,K}^x\to \hat s_{i,\bar K}^x, \hat s_{i,\bar K}^x\to \hat s_{i,K}^x, \hat s_{i,K}^z\to-\hat s_{i,\bar K}^z, \hat s_{i,\bar K}^z\to-\hat s_{i,K}^z.
\end{equation}
Therefore, $\hat H_{1/2}$ is invariant under $\hat{\mathcal P}$. This symmetry implies
\begin{equation}
\langle \hat S_K^x\rangle=\langle \hat S_{\bar K}^x\rangle,\qquad
\langle \hat S_K^z\rangle=-\langle \hat S_{\bar K}^z\rangle,
\label{eq:HM_valley_relation}
\end{equation}
at all temperatures. We therefore discuss only $\langle \hat S_K^x\rangle$ and $\langle \hat S_K^z\rangle$. The corresponding $\bar K$ components follow from \cref{eq:HM_valley_relation}.

The spin models in \crefrange{eq:H_QM_main}{eq:H_HM_main} retain spin and valley degrees of freedom while neglecting charge fluctuations. This is a nontrivial simplification, even in regimes where collective modes are spectrally isolated from the Stoner continuum, but it enables a numerically exact quantum Monte Carlo treatment of the resulting spin models.

\section{Stochastic-Series-Expansion Quantum Monte Carlo Method}
\label{sec:QMC}

We study \crefrange{eq:H_QM_main}{eq:H_HM_main} using stochastic-series-expansion quantum Monte Carlo \cite{Sandvik_2010}. We employ diagonal and directed-loop updates, supplemented by an open-walk update when the transverse field breaks spin conservation~\cite{Sandvik_2010,PhysRevE.66.046701,henelius2000monte,henelius2002quantum}. The bond decompositions, update procedures, and derivations of the estimators are given in \cref{appendix:sse_primer}. Here we briefly summarize the stochastic-series-expansion representation and define the observables used in the main text.

In stochastic-series-expansion quantum Monte Carlo, the Hamiltonian is written as $\hat H=-\sum_{a,b}\hat H_{a,b}+E_0$, where $E_0$ is an additive constant, and the partition function is expanded as a power series,
\begin{equation}
Z=
\sum_{\alpha}\sum_{n=0}^{\infty}
\frac{\beta^n}{n!}
\sum_{S_n}
\langle\alpha|
\prod_{p=1}^{n} \hat H_{a_p,b_p}
|\alpha\rangle,
\label{eq:SSE_Z}
\end{equation}
where $\beta=1/T$, $|\alpha\rangle$ is a basis state in the $z$ basis, and $S_n={(a_p,b_p)}_{p=1}^{n}$ denotes a sequence of local Hamiltonian operators, with $a_p$ labeling the operator type and $b_p$ the bond or site on which it acts. A Monte Carlo configuration therefore consists of a basis state and an operator string. Successively applying the operators generates propagated basis states, from which diagonal observables can be measured directly, while operators appearing explicitly in the Hamiltonian can also be estimated from their occurrence counts in the operator string.

\paragraph{Quarter-metal observables.}
For the quarter-metal phase, each propagated basis state specifies the eigenvalues $s_i^z,\tau_i^z=\pm1$ of $\hat s_i^z$ and $\hat\tau_i^z$. We define the spin and valley polarizations as
\begin{equation}
\hat S^z=\frac{1}{N}\sum_i \hat s_i^z,\qquad 
\hat T^z=\frac{1}{N}\sum_i \hat\tau_i^z.
\label{eq:QM_polarizations}
\end{equation}
We also define the local spin-valley locking
\begin{equation}
\hat C^{zz}=\frac{1}{N}\sum_i \hat s_i^z\hat\tau_i^z.
\label{eq:QM_composite}
\end{equation}
These quantities are evaluated directly from the propagated $z$-basis states and averaged over the Monte Carlo ensemble. On a finite lattice the signed averages $\langle\hat S^z\rangle$ and $\langle\hat T^z\rangle$ vanish by symmetry, so we report the root mean square order parameters, $\langle(\hat S^z)^2\rangle^{1/2}$ and $\langle(\hat T^z)^2\rangle^{1/2}$, computed from the second moments, the standard finite-size diagnostics for spontaneously broken order~\cite{Sandvik_2010}. 

\paragraph{Half-metal observables.}
We define the valley-resolved spin polarizations as 
\begin{equation}
\hat S_K^\mu=\frac{1}{N}\sum_i \hat s_{i,K}^\mu,\qquad
\hat S_{\bar K}^\mu=\frac{1}{N}\sum_i \hat s_{i,\bar K}^\mu, 
\end{equation}
where $\mu=x,y,z$. 
The valley-resolved out-of-plane polarization $\langle \hat S_\nu^z\rangle$ is measured directly from the propagated basis states. For $h_x\ne0$, the in-plane polarization is obtained from the transverse-field vertices using the stochastic-series-expansion counting estimator
\begin{equation}
\langle \hat S_\nu^x\rangle=\frac{\langle N_x^\nu\rangle}{N\beta h_x},
\qquad \nu\in\{K,\bar K\},
\label{eq:SSE_Skx}
\end{equation}
where $N_x^\nu$ counts transverse-field vertices for valley $\nu$ in the sampled operator string. The derivation and implementation of this estimator are given in \cref{appendix:sse_primer}. By \cref{eq:HM_valley_relation}, $\langle \hat S_K^x\rangle$ is the common in-plane polarization, while $\langle \hat S_K^z\rangle$ is the valley-staggered out-of-plane polarization. At $h_x=0$, we also measure the helicity modulus from the winding-number fluctuations,
\begin{equation}
\Upsilon=\frac{T}{2}\langle\bm W^2\rangle,
\label{eq:helicity}
\end{equation}
where $\bm W=(W_x,W_y)$ and
\begin{equation}
W_\mu=\frac{N_\mu^+-N_\mu^-}{L},
\qquad \mu\in\{x,y\}.
\label{eq:HM_winding}
\end{equation}
Here $N_\mu^\pm$ count spin-flip hopping vertices in the positive and negative $\mu$ directions. Onsite intervalley and transverse-field vertices do not transport spin between lattice sites and therefore do not contribute to $\bm W$. Further details of the estimators are given in \cref{appendix:sse_primer}.

\begin{figure}[t]
\centering
\includegraphics[width=0.9\linewidth]{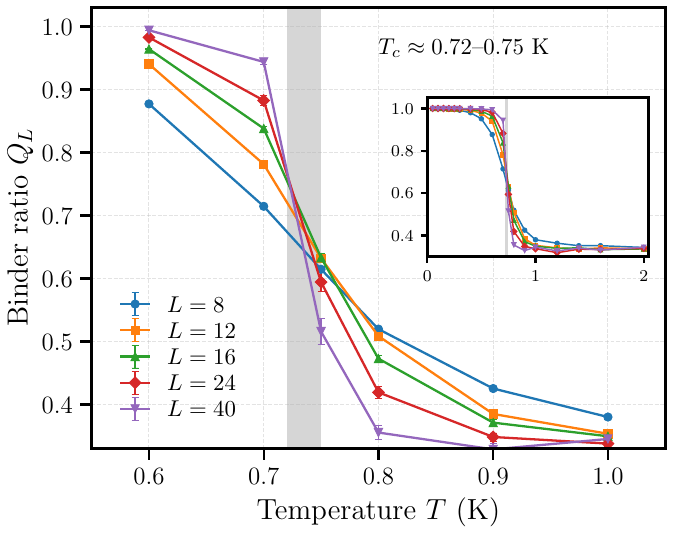}
\caption{Binder ratio $Q_L=\langle(\hat S^z)^2\rangle_L^2/\langle(\hat S^z)^4\rangle_L$ of the quarter-metal spin polarization for $L=8,12,16,24,$ and $40$, zoomed on the crossing region; the inset shows the full temperature range. The gray band indicates the larger-size crossing estimate $T_c\approx0.72$--$0.75~\mathrm{K}$.}
\label{fig:qm_binder}
\end{figure}

\begin{figure*}[t]
  \centering
  \includegraphics[width=1\linewidth]{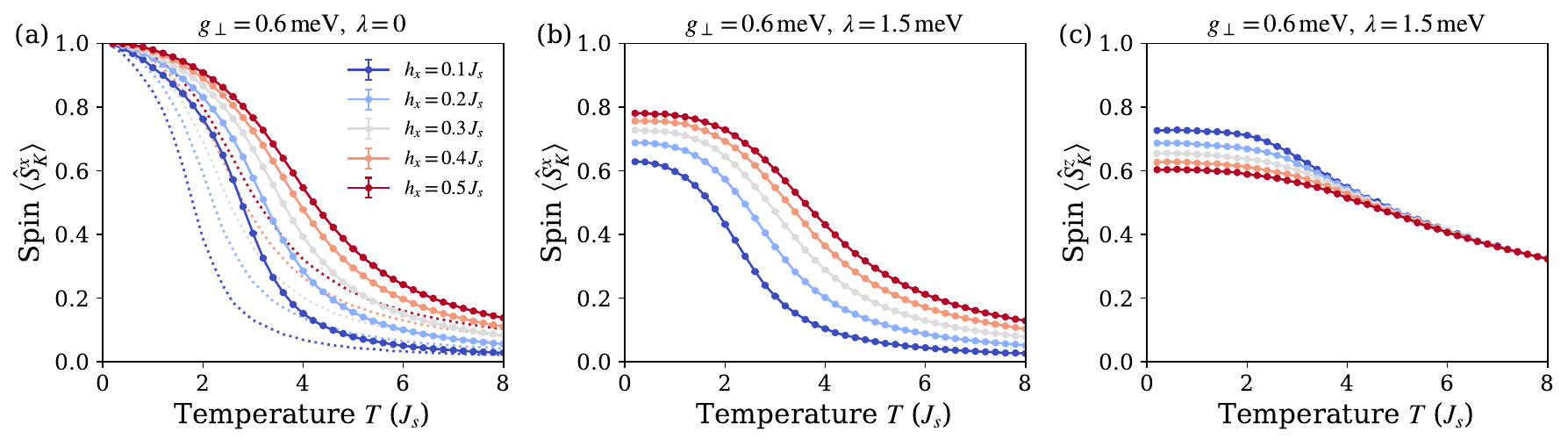}
\caption{%
Temperature dependence of the valley-resolved spin polarization in the half-metal phase on an $L=16$ square lattice with $J_s=0.30~\mathrm{meV}$. (a) In-plane polarization $\langle \hat S_K^x\rangle$ for bare rhombohedral trilayer graphene with $g_\perp=0.60~\mathrm{meV}$ and $\lambda=0$ at transverse fields $h_x/J_s=0.1$--$0.5$. Solid curves show the coupled-valley model, while dotted curves show the decoupled reference with $g_\perp=0$. For $\lambda=0$, $\langle \hat S_K^z\rangle=0$ by symmetry and is not shown. (b,c) In-plane and out-of-plane polarizations $\langle \hat S_K^x\rangle$ and $\langle \hat S_K^z\rangle$, respectively, for $g_\perp=0.60~\mathrm{meV}$ and $\lambda=1.50~\mathrm{meV}$ at the same transverse fields. The in-plane component is more strongly suppressed by increasing temperature than the SOC-favored out-of-plane component.
}
  \label{fig:spin_vector}
\end{figure*}

\begin{figure}[t]
  \centering
  \includegraphics[width=0.95\linewidth]{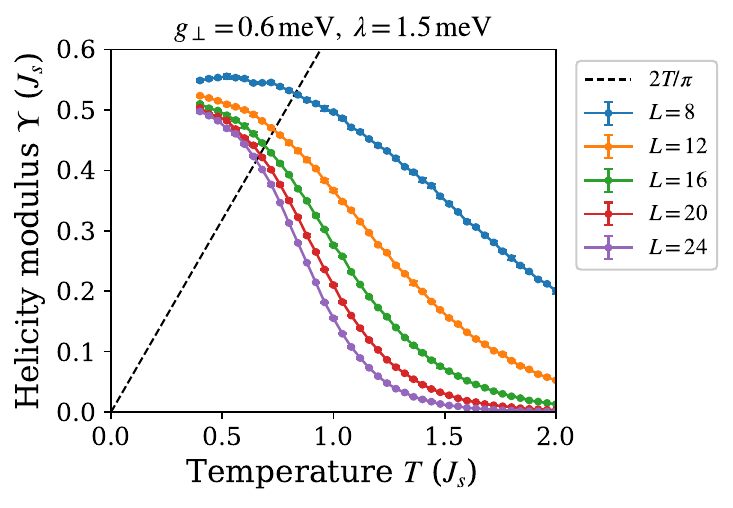}
  \caption{Helicity modulus $\Upsilon/J_s$ as a function of $T/J_s$ for $ J_s=0.30~\mathrm{meV}$, $g_\perp=0.60~\mathrm{meV}$, $\lambda=1.50~\mathrm{meV}$, and $h_x=0$ at lattice sizes $L=8,12,16,20,$ and $24$. The dashed line denotes the Nelson--Kosterlitz universal-jump condition $\Upsilon=2T/\pi$. The size-dependent intersections illustrate the finite-size drift toward the Berezinskii--Kosterlitz--Thouless transition. The thermodynamic transition temperature is obtained from the two-size analysis described in \cref{appendix:fss}.}
  \label{fig:helicity}
\end{figure}

\section{Results}
\label{sec:results}

\paragraph{Quarter-metal phase.} 
\Cref{fig:MF_vs_QMC}(b) shows the root mean square spin polarization $\langle(\hat S^z)^2\rangle^{1/2}$, RMS valley polarization $\langle(\hat T^z)^2\rangle^{1/2}$, and local spin-valley locking $\langle\hat C^{zz}\rangle$. The spin and valley polarizations increase continuously below $T_{\mathrm c}\approx 0.7~\mathrm{K}$.
We locate the transition quantitatively from the Binder ratio of the spin polarization, $Q_L=\langle(\hat S^z)^2\rangle_L^2/\langle(\hat S^z)^4\rangle_L$, whose crossings for different system sizes concentrate at $T_c\approx0.72$--$0.75~\mathrm{K}$ (\cref{fig:qm_binder}). The finite-size analysis and the assignment to the 2D Ising universality class are detailed in \cref{appendix:fss}.
Near $T_c$, individual Monte Carlo trajectories tunnel between the  low-energy doublets favored by spin-orbit coupling, namely, $(S^z, T^z) = (+1, +1)$ and  $(S^z, T^z)=(-1, -1)$.
This leads to strong fluctuations in the signed spin and valley polarizations, as shown in \cref{appendix:fss}. In contrast, $\hat C^{zz}$ has no global tunneling events. 
Unless stated otherwise, these data are generated with $J_s=0.10~\mathrm{meV}$, $J_\tau=0.15~\mathrm{meV}$, and $\lambda=0.05~\mathrm{meV}$. These parameters are obtained from self-consistent Hartree--Fock calculations for rhombohedral trilayer graphene as described in \cref{appendix:parameter_estimation}.

\paragraph{Half-metal phase.}
For the half-metal phase, the finite temperature property is very different for the bare rhombohedral graphene than for the 
TMD proximitized rhombohedral graphene. We first consider bare rhombohedral trilayer graphene, for which the small intrinsic spin--orbit scale ($\lambda\approx50~\mu\mathrm{eV}$) is neglected by setting $\lambda=0$, and then the proximity-enhanced regime with $\lambda=1.50~\mathrm{meV}$ relevant to graphene--TMD heterostructures~\cite{PhysRevB.92.155403,PhysRevB.99.245402,patterson2025superconductivity}.
For bare rhombohedral trilayer graphene, at $h_x=0$ and $\lambda = 0$ , the system is equivalent to two copies of the Heisenberg model coupled via $g_\perp$. The coupled Heisenberg model 
has a continuous $\mathrm{SU}(2)$ (total) spin-rotation symmetry. Therefore it has no spontaneous magnetic order at finite temperature in 2D by the Mermin--Wagner theorem. \cref{fig:spin_vector}(a) shows the in-plane spin-polarization induced by a finite transverse field $h_x$ for $\lambda=0$.  The dotted curves correspond to the decoupled limit, $g_\perp=0$, while the solid curves show the results for $g_\perp=0.60$~meV, which approximates the estimate $g_\perp\approx0.67$~meV in \cref{appendix:parameter_estimation}. Comparing these results shows that finite intervalley coupling enhances the field-induced magnetization for all temperatures.

Next, we consider spin--orbit-proximitized rhombohedral trilayer graphene with $g_\perp=0.60~\mathrm{meV}$ and $\lambda=1.50~\mathrm{meV}$. \cref{fig:spin_vector}(b,c) shows the corresponding valley-resolved polarizations $\langle \hat S_K^x\rangle$ and $\langle \hat S_K^z\rangle$ as functions of temperature and transverse field. With increasing temperature, the in-plane canting component $\langle \hat S_K^x\rangle$ is suppressed more rapidly than the out-of-plane component $\langle \hat S_K^z\rangle$.  This is because the out-of-plane polarization has a contribution coming from the single-particle term (spin--orbit interaction), whereas the spontaneous in-plane polarization arises purely from interactions. At finite $h_x$, the in-plane polarization also contains a field-induced contribution. The polarization in the opposite valley follows from the symmetry operation in \cref{eq:HM_hidden_symmetry}, see \cref{eq:HM_valley_relation}.
Because the transverse field couples linearly to the common in-plane component, it affects $\langle \hat S_K^x\rangle$ more strongly than $\langle \hat S_K^z\rangle$, as seen in \cref{fig:spin_vector}(b,c). Results for other values of $g_\perp$ are presented in \cref{appendix:hm_g_dependence}.

At zero transverse field, the Hamiltonian has a global $\mathrm{U}(1)$ symmetry under spin rotations about the $z$ axis, allowing the canted half-metal phase to undergo a finite-temperature Berezinskii--Kosterlitz--Thouless transition. We locate this transition using the Nelson--Kosterlitz universal-jump condition, $\Upsilon(T_{\rm KT})=2k_{\rm B}T_{\rm KT}/\pi$~\cite{nelson1977universal}. \Cref{fig:helicity} shows the helicity modulus for $g_\perp=0.60~\mathrm{meV}$ and $\lambda=1.50~\mathrm{meV}$. Its intersections with the universal-jump line exhibit the systematic finite-size drift characteristic of a Berezinskii--Kosterlitz--Thouless transition. Applying the corrected two-size procedure described in \cref{appendix:fss}, we obtain the thermodynamic-limit estimate $k_{\rm B}T_{\rm KT}/J_s=0.5666\pm0.0008$, corresponding to $T_{\rm KT}=1.973\pm0.003~\mathrm{K}$ for $J_s=0.30~\mathrm{meV}$.

\section{Summary}
\label{sec:summary}

We use quantum Monte Carlo simulations to study finite-temperature transverse spin--valley fluctuations beyond the routine Hartree--Fock and time-dependent Hartree--Fock approximations. These fluctuations are described by an effective local-moment model whose parameters are extracted from constrained Hartree--Fock calculations. Perhaps most surprising is that, despite the seemingly ad hoc nature of the local-moment description, this approach yields a transition temperature $T_c\approx0.7$~K, comparable to the experimental scale~\cite{Holleis_2025}, without any fitting.

For the half-metal phase, we study the field-induced magnetization both with and without proximity-enhanced spin--orbit coupling. When spin--orbit coupling is negligible, intervalley Hund's coupling enhances the field-induced magnetic response. When spin--orbit coupling is large (e.g., induced by a neighboring transition-metal dichalcogenide), the ground state is the familiar spin canted phase. When temperature increases, the in-plane polarization decreases more rapidly than the out-of-plane polarization, since the latter has a single-particle (spin--orbit coupling) contribution. These results show that the Monte Carlo simulations of an effective local-moment model can capture some important properties of the finite-temperature magnetic response despite the system's itinerant character. Incorporating charge dynamics is a natural next step toward a more complete description.

\bigskip 
\paragraph*{Acknowledgments.}
We thank Ribhu Kaul for helpful discussions regarding Monte Carlo algorithms.  
This work was supported by the U.S. Department of Energy, Office of Science, Office of Basic Energy Sciences under Award Number DE-SC0024346. We are grateful to the University of Kentucky Center for Computational Sciences and Information Technology Services Research Computing for their support and use of the Morgan Compute Cluster and associated research computing resources.

\appendix
\crefalias{section}{appendix}
\section{Stochastic-series-expansion quantum Monte Carlo: bond decomposition, updates, and estimators}
\label{appendix:sse_primer}

This appendix gives the bond and vertex decompositions used in the stochastic-series-expansion quantum Monte Carlo simulations, together with the update procedures and estimators specific to the present models. General background on the high-temperature expansion, the no-branching basis, and directed-loop updates can be found in Refs.~\cite{Sandvik_2010,henelius2000monte,henelius2002quantum,PhysRevE.66.046701}.

Both models are defined on the same $L\times L$ square lattice with periodic boundary conditions. There are $N_b=2N$ nearest-neighbor bonds, with each bond counted once. Each bond $b$ connects sites $i(b)$ and $j(b)$, and the coordination number is $z=4$. Onsite terms are rewritten as bond sums using
\begin{equation}
\sum_i f_i=\frac{1}{z}\sum_b\left[f_{i(b)}+f_{j(b)}\right].
\end{equation}
Throughout this appendix, the local spin and valley operators use the saturation-1 convention: $\hat s^z$, $\hat\tau^z$, and the valley-resolved $\hat s^z_{K},\hat s^z_{\bar K}$ have eigenvalues $\pm1$, and
\begin{equation}
\hat s^\pm=\frac{\hat s^x\pm i \hat s^y}{2},
\end{equation}
with analogous definitions for $\hat\tau^\pm$ and $\hat s^\pm_{K},\hat s^\pm_{\bar K}$. Site indices are suppressed where unambiguous.

\paragraph{Quarter metal.}
The quarter-metal Hamiltonian is decomposed into diagonal and off-diagonal bond operators as
\begin{align}
\hat H_{1/4}&=-\sum_b\left(\hat h_{D,b}+\hat h^s_{O,b}+\hat h^\tau_{O,b}\right)+(J_s+J_\tau)N_b+\frac{\lambda N_b}{z}.
\label{eq:H_quarter_metal}
\end{align}
The additive constant is chosen so that all diagonal matrix elements of $\hat h_{D,b}$ are non-negative. The bond operators are
\begin{subequations}
\label{eq:bond_ops}
\begin{align}
\hat h_{D,b}&=
J_s\left(1+\hat s^z_{i(b)}\hat s^z_{j(b)}\right)+J_\tau\left(1+\hat\tau^z_{i(b)}\hat\tau^z_{j(b)}\right)
\nonumber\\
&\quad+\frac{\lambda}{2z}\left(2+\hat s^z_{i(b)}\hat\tau^z_{i(b)}+\hat s^z_{j(b)}\hat\tau^z_{j(b)}\right),\\
\hat h^s_{O,b}&=
2J_s\left(\hat s^+_{i(b)}\hat s^-_{j(b)}+\hat s^-_{i(b)}\hat s^+_{j(b)}\right),\\
\hat h^\tau_{O,b}&=2J_\tau\left(\hat\tau^+_{i(b)}\hat\tau^-_{j(b)}+\hat\tau^-_{i(b)}\hat\tau^+_{j(b)}\right).
\end{align}
\end{subequations}

For the quarter-metal model, the no-branching condition gives 16 diagonal vertices, 8 spin-exchange vertices, and 8 valley-exchange vertices.

\paragraph{Half metal.}
The half-metal model uses the four-state onsite basis
\begin{equation}
\left\{|s_K^z,s_{\bar K}^z\rangle\right\},\qquad
s_K^z,s_{\bar K}^z=\pm 1.
\end{equation}
Using the bond representation of the onsite terms introduced above, the Hamiltonian becomes
\begin{equation}
\hat H_{1/2}=-\sum_b\left(\hat h_{D,b}+\hat h_{O,b}\right)
+2J_sN_b+\frac{2g_\perp N_b}{z}+\frac{2\lambda N_b}{z}.
\label{eq:H_half_metal}
\end{equation}
The additive constants make the diagonal bond matrix elements non-negative for $J_s,g_\perp,\lambda\ge0$. 
The diagonal bond operator is
\begin{equation}
\label{eq:HM_bond_op_D}
\begin{aligned}
\hat h_{D,b}&=J_s\left(\hat s^z_{i(b),K}\hat s^z_{j(b),K}+\hat s^z_{i(b),\bar K}\hat s^z_{j(b),\bar K}+2\right)\\
&\quad+\frac{g_\perp}{z}\left(\hat s^z_{i(b),K}\hat s^z_{i(b),\bar K}+\hat s^z_{j(b),K}\hat s^z_{j(b),\bar K}+2\right)\\
&\quad+\frac{\lambda}{2z}\left(\hat s^z_{i(b),K}+\hat s^z_{j(b),K}-\hat s^z_{i(b),\bar K}-\hat s^z_{j(b),\bar K}+4\right).
\end{aligned}
\end{equation}
The off-diagonal bond operator contains intravalley exchange, intervalley exchange, and transverse-field vertices,
\begin{subequations}
\label{eq:HM_bond_ops_O}
\begin{align}
\hat h_{O,b}&=\sum_{\nu,\eta=\pm}\hat h_{O,b}^{\nu\eta}+\sum_m \hat h_{O,b}^{g_\perp,m}+\sum_{\nu,m}\hat h_{O,b}^{h_x,\nu,m},\\
\hat h_{O,b}^{K\pm}&=2J_s \hat s^\pm_{i(b),K}\hat s^\mp_{j(b),K},\\
\hat h_{O,b}^{\bar K\pm}&=2J_s \hat s^\pm_{i(b),\bar K}\hat s^\mp_{j(b),\bar K},\\
\hat h_{O,b}^{g_\perp,m}&=\frac{2g_\perp}{z}\left(\hat s_{m,K}^+\hat s_{m,\bar K}^- +\hat s_{m,K}^-\hat s_{m,\bar K}^+\right),\\
\hat h_{O,b}^{h_x,K,m}&=\frac{h_x}{z}\left(\hat s_{m,K}^++\hat s_{m,K}^-\right),\\
\hat h_{O,b}^{h_x,\bar K,m}&=\frac{h_x}{z}\left(\hat s_{m,\bar K}^++\hat s_{m,\bar K}^-\right).
\end{align}
\end{subequations}
Here $\nu\in\{K,\bar K\}$ and $m\in\{i(b),j(b)\}$.
Each off-diagonal operator maps a basis state to a single new basis state, preserving the no-branching structure required by the stochastic-series-expansion representation. The transverse-field terms generate single-leg vertices. Directed-loop updates act within the two valley sectors, while the intervalley vertices allow spin exchange between them at a given site.

\paragraph{Updates.}
Each Monte Carlo sweep contains diagonal updates, free-spin flips, and directed-loop updates. Let $n$ denote the number of nonidentity operators in an operator string of cutoff $M$. Diagonal operators are inserted using the Metropolis ratio
\begin{equation}
\frac{\beta N_b\langle\alpha|\hat h_{D,b}|\alpha\rangle}{M-n}.
\end{equation}
The reciprocal ratio is used for removal. Spins at sites untouched by the operator string can be flipped freely because such flips leave the weight of the Monte Carlo configuration unchanged. Directed loops propagate a defect through the linked space-time graph using exit-leg probabilities that satisfy local detailed balance,
\begin{equation}
P(x|e)\omega(\eta)=P(e|x)\omega(\eta'),
\end{equation}
where $\eta$ and $\eta'$ denote the vertex states before and after the local update~\cite{PhysRevE.66.046701}. The truncation order $M$ is adjusted dynamically to maintain $\langle n\rangle/M<0.85$. Statistical uncertainties are estimated using binning analysis~\cite{Flyvbjerg1989}.

For $h_x\ne0$, the transverse-field vertices break conservation of the total spin projection. We therefore use the open-walk directed-loop update of Henelius et al.~\cite{henelius2002quantum}. The entrance leg is not flipped at the beginning of the walk, allowing a nonconserving vertex to be created or removed. The walk can terminate at a later compatible vertex, which preserves the periodic boundary condition along the operator string.

\paragraph{Estimators.}
For an operator family that appears explicitly in the stochastic-series-expansion decomposition, its expectation value can be obtained from the counting identity
\begin{equation}
\langle \hat h_j\rangle
=
\frac{\langle n_j\rangle}{\beta},
\end{equation}
where $n_j$ is the number of occurrences of that operator family in the sampled operator string~\cite{Sandvik_2010}. Applying this identity to the transverse-field vertices gives the estimator for $\hat S_\nu^x$ in \cref{eq:SSE_Skx}. Applying the same counting procedure to the spatial spin-exchange vertices gives the winding numbers in \cref{eq:HM_winding}. 

The quarter-metal polarizations are measured directly from the propagated basis states according to \cref{eq:QM_polarizations}; the reported RMS values $\langle(\hat S^z)^2\rangle^{1/2}$ and $\langle(\hat T^z)^2\rangle^{1/2}$ are computed from the accumulated second moments of the spatially averaged polarizations. The local spin-valley locking is measured as the signed average $\langle\hat C^{zz}\rangle$; since $\hat C^{zz}$ is invariant under the joint $\mathbb Z_2$ transformation, this estimator is insensitive to global flips between the two symmetry-related sectors and requires no symmetrization.

\paragraph{Implementation.}
The Monte Carlo configuration is represented by an operator string together with the linked vertex graph of Ref.~\cite{Sandvik_2010}. Directed-loop and open-walk updates are performed on this graph, after which the propagated site states and operator string are reconstructed for measurements.

\section{Estimating parameters from mean-field methods}
\label{appendix:parameter_estimation}
This appendix details the estimation of energy anisotropies and stiffness constants for the quarter-metal and half-metal phases of dual-gated, encapsulated rhombohedral trilayer graphene. These parameters serve as inputs for the quantum Monte Carlo simulations in the main text. Throughout the paper, the spin and valley operators are Pauli-vector operators in the saturation-1 convention, $|\langle\hat{\vec s}\rangle|=|\langle\hat{\vec\tau}\rangle|=1$ for a fully polarized state. The vector $\hat{\vec\sigma}$ denotes the Pauli operator in the layer-resolved sublattice space.

We solve the mean-field equations self-consistently using a constrained Fock approximation. The mean-field Bloch Hamiltonian reads
\begin{align}
     \mathcal{\hat H}^{\text{MF}}_{\vec{k}} = \mathcal{\hat H}_{0,\vec{k}} + \hat\Sigma^{\text{F}}_{\vec{k}},\label{eq:mean-field_hamiltonian}
\end{align}
where $\mathcal{\hat H}_{0,\vec{k}}$ is the kinetic single-particle Hamiltonian with Slonczewski--Weiss--McClure parameters. The Fock self-energy $\hat\Sigma^{\text{F}}_{\vec{k}}$ is built from the density matrix $\hat\rho_{\vec{k}}$ and the dual-gate-screened Coulomb interaction $V_{\vec{q}}$,
\begin{align}
    \hat\Sigma^{\text{F}}_{\vec{k}} = -\frac{1}{A}\sum_{\vec{k}'} V_{\vec{k}-\vec{k}'}\, \hat\rho_{\vec{k}'},
    &&
    V_{\vec{q}} = \frac{2\pi e^2}{\varepsilon_r\, |\vec{q}|}\, \tanh(|\vec{q}|\, d),\label{eq:Fock}
\end{align}
where $d$ is the gate distance, $A$ is the sample area and $\varepsilon_r$ is an effective dielectric constant. We take $d = 50$~nm and $\varepsilon_r = 15$; see our earlier work~\cite{PhysRevB.109.L060409} for details.
The density matrix depends on the eigenstates $\ket{\psi_{n,\vec{k}}}$ and eigenvalues $\varepsilon_{n,\vec{k}}$ of $\mathcal{\hat H}^{\text{MF}}_{\vec{k}}$, forming a self-consistent loop that we iterate to convergence:
\begin{align}
    \hat\rho_{\vec{k}} = \sum_n n_{\rm F}(\varepsilon_{n,\vec{k}})\, \ket{\psi_{n,\vec{k}}}\bra{\psi_{n,\vec{k}}}.\label{eq:rho_mf}
\end{align}
The total energy density is then
\begin{align}
    \mathcal{E} = \int \frac{d^2k}{(2\pi)^2}\, \operatorname{Tr}\!\left[\left(\mathcal{\hat H}_{0,\vec{k}} + \tfrac{1}{2}\,\hat\Sigma^{\text{F}}_{\vec{k}}\right) \hat\rho_{\vec{k}}\right].
\end{align}
The low-energy electrons carry four isospin flavors ($K\!\uparrow$, $K\!\downarrow$, $\bar{K}\!\uparrow$, $\bar{K}\!\downarrow$), and spontaneous symmetry breaking yields an unequal occupation $\vec{\nu} = (\nu_{K\uparrow}, \nu_{K\downarrow}, \nu_{\bar{K}\uparrow}, \nu_{\bar{K}\downarrow})$ with $\sum_f \nu_f = 1$. Our key calculated parameters are summarized in \cref{tab:hm_qm_table}.

\begin{table}[b]
\centering
\caption{Model parameters for the quarter-metal and half-metal phases of rhombohedral trilayer graphene at $U_D = 36$ meV, with $n_e = -3.7 \times 10^{11}~\mathrm{cm}^{-2}$ (quarter-metal) and $n_e = -1.0 \times 10^{12}~\mathrm{cm}^{-2}$ (half-metal). The Heisenberg-bond couplings $J_s, J_\tau$ are obtained from the continuum HF stiffness via the matching procedure of \cref{appendix:parameter_estimation}; the intervalley exchange $g_\perp$ follows from the projection of the intervalley Hund interaction onto the valley-resolved spin polarizations discussed below. The spin-model couplings are used in the quantum Monte Carlo simulations of \cref{sec:results}.}
\label{tab:hm_qm_table}
\begin{tabular}{@{}lcc@{}}
\hline
\hline
\textbf{Parameter} & \textbf{QM} & \textbf{HM} \\
\hline
Order vector $\vec{\nu}$ & $(1,0,0,0)$ & $(1/2,0,1/2,0)$\\
Spin polarization $|\braket{\hat S^z}|$ & 1 & 1\\
Valley polarization $|\braket{\hat T^z}|$ & 1 & 0\\
Spin coupling $J_s$ (meV) & 0.1 & 0.3\\
Valley coupling $J_\tau$ (meV) & 0.15 & --\\
Intervalley exchange $g_\perp$ (meV) & -- & 0.67\\
\hline
\hline
\end{tabular}
\end{table}

\subsection{Intervalley Hund's coupling}

To incorporate the intervalley Hund's scattering process, we add the following contribution to the mean-field Hamiltonian,
\begin{align}
  \hat\Sigma^{\rm IV}
  &= g_{\rm IV}\sum_{\substack{l=1,2,3\\ \sigma=A,B}}\sum_{a=x,y}
      \left[
        \operatorname{Tr}\!\left(\hat\tau_a \hat P^l_\sigma\hat{\bar\rho}\right)
        \hat \tau_a\hat P^l_\sigma
        - \hat\tau_a \hat P^l_\sigma\hat{\bar\rho}\,
        \hat\tau_a\hat P^l_\sigma
      \right], \nonumber\\
  \hat{\bar\rho}
  &=\int\frac{d^2k}{(2\pi)^2}\hat\rho_{\vec k},
  \label{eq:ivcs_self_energy}
\end{align}
where $\hat P^l_\sigma$ projects onto layer $l$ and sublattice
$\sigma$. This interaction reduces the independent
$\mathrm{SU}(2)_K\times\mathrm{SU}(2)_{\bar K}$ spin-rotation
symmetry to the global spin $\mathrm{SU}(2)$ symmetry and, for
$g_{\rm IV}>0$, energetically favors parallel spin alignment between
the two valleys.

The corresponding contribution to the mean-field energy density is
\begin{align}
    \mathcal{E}^{\mathrm{IV}}
    =\frac{g_{\rm IV}}{2}
    \sum_{\substack{l=1,2,3\\ \sigma=A,B}}\sum_{a=x,y}
      \left[
        \operatorname{Tr}\!\left(
        \hat\tau_a \hat P^l_\sigma\hat{\bar\rho}\right)^2
        -\operatorname{Tr}\!\left(
        \hat\tau_a \hat P^l_\sigma\hat{\bar\rho}\,
        \hat\tau_a\hat P^l_\sigma\hat{\bar\rho}
        \right)
      \right].
\end{align}
To express this interaction in terms of the valley-resolved spin
polarizations, we write the density matrix in valley-block form,
\begin{align}
    \hat{\bar\rho}
    =
    \begin{pmatrix}
        \hat\rho_K & \hat\Delta\\
        \hat\Delta^\dagger & \hat\rho_{\bar K}
    \end{pmatrix}.
\end{align}
Throughout the half-metal phase considered here there is no
intervalley coherence, $\hat\Delta=0$. Consequently, the Hartree
contribution vanishes,
\begin{align}
    \sum_{a=x,y}
    \operatorname{Tr}\!\left(
    \hat\tau_a\hat P^l_\sigma\hat{\bar\rho}\right)^2=0,
\end{align}
while the exchange contribution becomes
\begin{align}
    \sum_{a=x,y}
    \operatorname{Tr}\!\left(
    \hat\tau_a\hat P^l_\sigma\hat{\bar\rho}\,
    \hat\tau_a\hat P^l_\sigma\hat{\bar\rho}
    \right)
    =
    4\operatorname{Tr}_s\!\left(
    \hat\rho^l_{K,\sigma}
    \hat\rho^l_{\bar K,\sigma}
    \right),
\end{align}
where
\begin{align}
    \hat\rho^l_{\tau,\sigma}
    \equiv
    \hat P^l_\sigma\hat\rho_\tau\hat P^l_\sigma
\end{align}
is the spin density matrix projected onto sublattice $\sigma$ of
layer $l$. The intervalley Hund's coupling energy therefore reduces to
\begin{align}
    \mathcal{E}^{\mathrm{IV}}
    =
    -2g_{\rm IV}
    \sum_{\substack{l=1,2,3\\ \sigma=A,B}}
    \operatorname{Tr}_s\!\left(
    \hat\rho^l_{K,\sigma}
    \hat\rho^l_{\bar K,\sigma}
    \right).
    \label{eq:IVCS_energy}
\end{align}

We now decompose each projected spin density matrix as
\begin{align}
    \hat\rho^l_{\tau,\sigma}
    =
    \frac{1}{2}
    \left(
    n^l_{\tau,\sigma}\hat{\mathrm I}
    +
    \vec n^l_{\tau,\sigma}\cdot\hat{\vec s}
    \right),
\end{align}
where $n^l_{\tau,\sigma}$ is the charge density and
$\vec n^l_{\tau,\sigma}$ is the spin-polarization density in the
corresponding valley, layer, and sublattice sector. Using
$\operatorname{Tr}_s(\hat{\mathrm I})=2$ and
$\operatorname{Tr}_s(\hat s^a \hat s^b)=2\delta_{ab}$ gives
\begin{align}
    \operatorname{Tr}_s\!\left(
    \hat\rho^l_{K,\sigma}
    \hat\rho^l_{\bar K,\sigma}
    \right)
    =
    \frac{1}{2}
    \left[
    n^l_{K,\sigma}n^l_{\bar K,\sigma}
    +
    \vec n^l_{K,\sigma}\cdot
    \vec n^l_{\bar K,\sigma}
    \right].
\end{align}
\Cref{eq:IVCS_energy} thus becomes
\begin{align}
    \mathcal{E}^{\mathrm{IV}}
    &=
    -g_{\rm IV}
    \sum_{\substack{l=1,2,3\\ \sigma=A,B}}
    \left[
    n^l_{K,\sigma}n^l_{\bar K,\sigma}
    +
    \vec n^l_{K,\sigma}\cdot
    \vec n^l_{\bar K,\sigma}
    \right]
    \nonumber\\
    &=
    -g_{\rm IV}
    \sum_{\substack{l=1,2,3\\ \sigma=A,B}}
    \vec n^l_{K,\sigma}\cdot
    \vec n^l_{\bar K,\sigma}
    +\mathrm{const.},
    \label{eq:hund_energy_1}
\end{align}
where the density-dependent term is independent of the relative spin
orientation and has therefore been absorbed into the constant.

In rhombohedral trilayer graphene, the low-energy wave functions are predominantly localized on
the outer-layer sublattices. Moreover, at sufficiently large
displacement field the carriers are strongly polarized toward one of
these outer-layer sublattices. We may therefore project the
valley-resolved spin density onto the dominant sublattice and write
\begin{align}
    \sum_{\substack{l=1,2,3\\ \sigma=A,B}}
    \vec n^l_{K,\sigma}\cdot
    \vec n^l_{\bar K,\sigma}
    \simeq
    \vec n_K\cdot\vec n_{\bar K}.
\end{align}
For the valley-balanced half metal, the two valleys each contain half
of the total carrier density. For fully spin-polarized valleys,
\begin{align}
    \vec n_K
    =\frac{|n_e|}{2}\vec S_K,
    \qquad
    \vec n_{\bar K}
    =\frac{|n_e|}{2}\vec S_{\bar K},
\end{align}
where $\vec S_\tau$ is a unit vector specifying the spin orientation
in valley $\tau$. The spin-dependent Hund energy density is therefore
\begin{align}
    \mathcal{E}^{\mathrm{IV}}
    &=
    -\frac{g_{\rm IV}n_e^2}{4}
    \vec S_K\cdot\vec S_{\bar K}
    +\mathrm{const.}
\end{align}
Dividing by the total carrier density and dropping the spin-independent
constant gives the Hund energy per carrier,
\begin{align}
    \frac{\mathcal{E}^{\mathrm{IV}}}{|n_e|}
    =
    -\frac{g_{\rm IV}|n_e|}{4}
    \vec S_K\cdot\vec S_{\bar K}.
    \label{eq:hund_per_carrier}
\end{align}
In the effective Hamiltonian of \cref{eq:H_HM_main}, each lattice site carries two valley-resolved Pauli spins, with \(\langle\hat{\vec s}_{i,\tau}\rangle=\vec S_\tau\) in a fully polarized product state. The Hund energy per lattice site is therefore
\begin{align}
    \frac{\langle\hat H_{\mathrm{Hund}}\rangle}{N}
=-g_\perp\,\vec S_K\cdot\vec S_{\bar K},
\end{align}

where \(N\) counts lattice sites. This normalization is per site, rather than per individual valley-resolved spin. We determine \(g_\perp\) by matching this energy per site to the microscopic Hund energy per total carrier in \cref{eq:hund_per_carrier}, obtaining
\begin{align}
    g_\perp=\frac{g_{\rm IV}|n_e|}{4}
    .
\end{align}

We use $g_{\rm IV}=269~\mathrm{meV\,nm^2}$ following
Ref.~\cite{PhysRevLett.134.046501}. The corresponding effective
spin-model coupling therefore scales linearly with carrier density,
$g_\perp=0.67$ meV for $n_e=10^{12}\rm cm^{-2}$. Its precise value can also acquire
corrections from the layer--sublattice distribution of the low-energy
wave functions, and hence can vary with both carrier density and
displacement field.

\begin{figure*}
    \centering
    \includegraphics[width=1\linewidth]{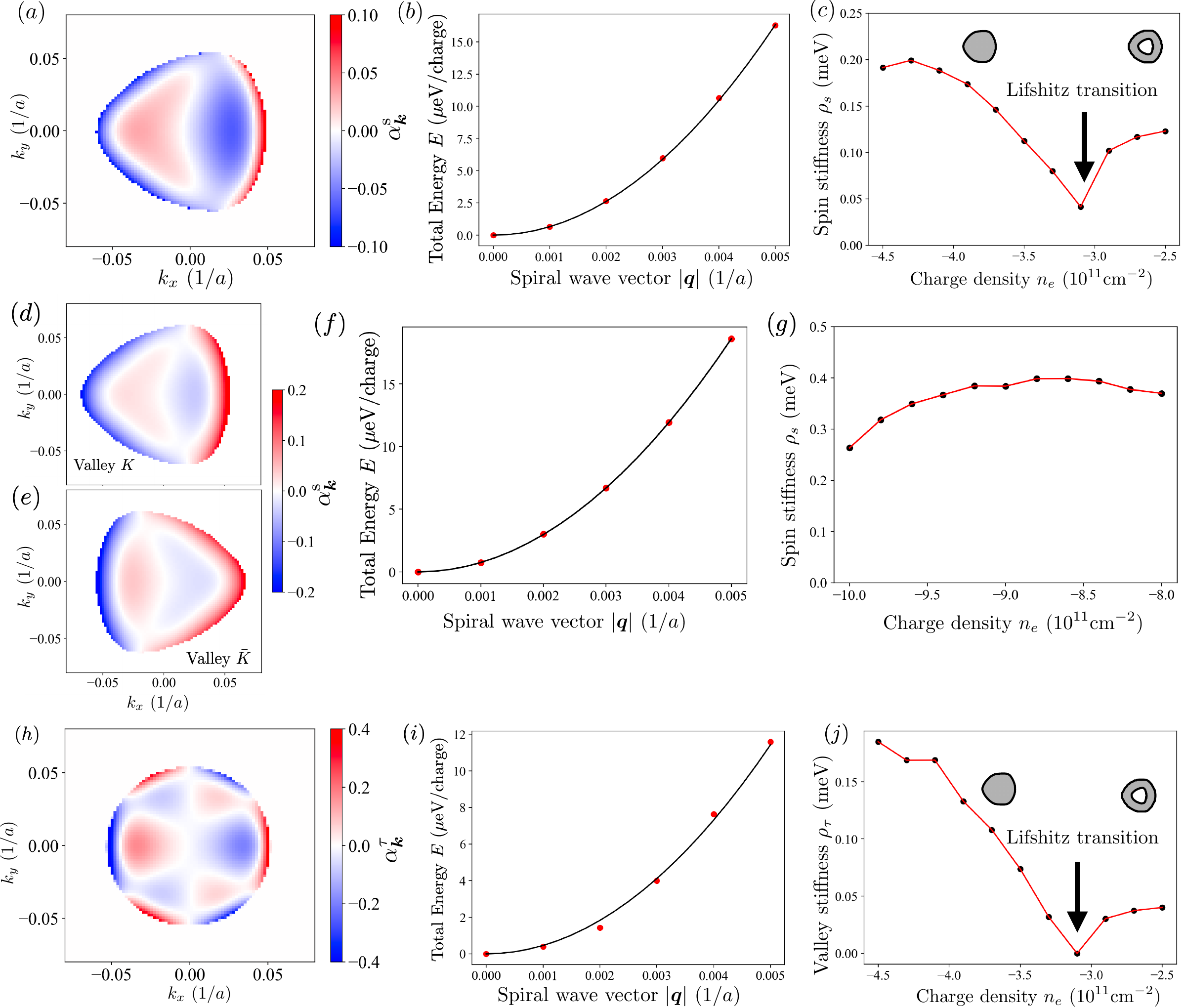}
    \caption{
    Spin and valley stiffnesses extracted from spiral states at $U_D = 36$ meV. (a) Momentum-resolved spin tilt angle $\alpha_{\vec k}^{s}$ for a spin spiral constructed from the QM state. (b) Spin-spiral energy cost per carrier relative to the uniform QM state versus spiral wave-vector magnitude $|\vec q|$; the black curve is a quadratic fit. (c) Spin stiffness $\rho_s$ versus carrier density. (d),(e) Momentum-resolved spin tilt angles in valleys $K$ and $\bar K$, respectively, for a spin spiral constructed from the HM state. (f),(g) Corresponding spiral energy cost and spin stiffness for the HM state. (h)–(j) Momentum-resolved valley-pseudospin tilt angle $\alpha_{\vec k}^{\tau}$, valley-spiral energy cost, and valley stiffness $\rho_\tau$ for the intervalley-coherent QM state. Black curves in (f) and (i) are quadratic fits. Arrows in (c) and (j) mark the annular Lifshitz transition, with schematic Fermi surfaces shown on either side.
    }
    \label{fig:spin_valley_stiffness}
\end{figure*}

\subsection{Spin and valley stiffness}

To extract the spin stiffness, we compute the energy cost of a long-wavelength spin spiral. A closely related self-consistent Hartree--Fock construction was previously used to evaluate the valley stiffness of an intervalley coherent quarter metal \cite{das2025momentum}. In that work, the spiral described a spatial winding of the intervalley coherent phase and was evaluated at a different carrier density and interlayer potential. Here, we apply the same general spiral construction to the flavor-polarized states relevant to the present work, namely, the spin-polarized, valley-Ising quarter metal and the spin-polarized half metal. The resulting textures and stiffnesses therefore characterize distinct Hartree--Fock ground states and parameter regimes. The spin spiral is implemented by applying opposite momentum shifts $\pm\vec q/2$ to the spin-up and spin-down sectors of the Bloch Hamiltonian,
\begin{align}
    \mathcal{\hat H}_{0,\vec k,\vec q}
    =
    \mathcal {\hat H}^{\uparrow}_{0,\vec k-\vec q/2}
    +
    \mathcal{\hat H}^{\downarrow}_{0,\vec k+\vec q/2},
\end{align}
where $\mathcal {\hat H}^{\uparrow,\downarrow}_{0,\vec k}$ denote the spin-projected blocks of the non-interacting Hamiltonian. The mean-field Hamiltonian
\begin{align}
    \mathcal {\hat H}^{\rm MF}_{\vec k,\vec q}
    =
    \mathcal {\hat H}_{0,\vec k,\vec q}
    +
    \hat \Sigma^{\rm F}_{\vec k},
\end{align}
is then solved self-consistently for each $\vec q$, with $\hat\Sigma^{\rm F}_{\vec k}$ the Fock self-energy defined in \cref{eq:Fock}.

The resulting momentum-resolved spin texture [\cref{fig:spin_valley_stiffness}] is characterized by the canting angle measured from XY plane
\begin{align}
    \tan \alpha^s_{\vec k}
    =
    \frac{\mathrm{Tr}(\hat \rho_{\vec k}\hat s_z)}
         {\mathrm{Tr}(\hat \rho_{\vec k}\hat s_x)},
\end{align}
where $\hat \rho_{\vec k}$ is the self-consistent density matrix. From the total energy density $\mathcal E_{\vec q}$ of the spiral state, we define the spin stiffness $\rho_s$ through the long-wavelength expansion
\begin{align}
    \label{eq:rho_s_def}
    \mathcal E_{\vec q}-\mathcal E_0
    =
    \frac{\rho_s}{2}\,|\vec q|^2
    +\mathcal O(q^4),
\end{align}
where $\mathcal E_0$ is the energy density of the uniform ground state. The quantity $\rho_s$ has units of energy. \cref{fig:spin_valley_stiffness} (b,f) show the change in energy vs. wave-vector $|\vec q|$ for QM and HM phase, respectively. From the quadratic fitting, we obtain $\rho_s=0.1$ meV in quarter-metal and $\rho_s=0.3$ meV in half-metal phase.

The continuum free-energy functional describing long-wavelength spin textures is
\begin{align}
    \delta F
    =
    \frac{\tilde \rho_s}{2S_0^2}
    \int d^2r
    |\nabla \vec S(\vec r)|^2,
    \label{eq:continuum_stiffness}
\end{align}
where $S_0=|\vec S(\vec r)|$ is the magnitude of the spin order parameter and $\delta F$ is measured relative to the uniform ground state. For a long-wavelength spin spiral with fixed magnitude, the order parameter takes the form
\begin{align}
    \vec S(\vec r)=S_0\left(
\cos(\vec q\cdot\vec r)\,\vec e_{x}
+
\sin(\vec q\cdot\vec r)\,\vec e_{z}
\right).
\end{align}
Substituting the equation above into \cref{eq:continuum_stiffness}, we get
\begin{align}
    \frac{\delta F}{A}
    =
    \frac{\tilde \rho_s}{2}
    |\vec q|^2
\end{align}
for an isotropic spin system. Comparison with \cref{eq:rho_s_def} shows that the coefficient entering the continuum free-energy functional is identical to the stiffness extracted from the spin-spiral calculation, $\tilde\rho_s=\rho_s$. 

The static stiffness extracted above is also related to the long-wavelength collective modes obtained within time-dependent Hartree--Fock theory. In a spin-polarized state with exact spin-rotation symmetry, the ferromagnetic Goldstone mode has the long-wavelength form
\begin{equation}
\hbar\omega(\vec q)=\mathcal D|\vec q|^2+\mathcal O(q^4),
\end{equation}
where the mode curvature $\mathcal D \sim \rho_s/M$ is determined by the static spin stiffness together with the Berry-phase coefficient, i.e., the density $M$ of the polarized moment. Thus, after fixing this dynamical normalization, the time-dependent Hartree--Fock spin-wave dispersion provides an independent route to $\rho_s$. Here we instead determine $\rho_s$ directly from the static spin-spiral energy, since it maps directly onto the exchange coupling of the local-moment Hamiltonian. The time-dependent Hartree--Fock spectra in \cref{fig:R2G_R3G_TDHF} are used primarily to establish that the relevant flavor-rotation modes form well-defined low-energy branches separated from the Stoner continuum over a substantial range of momenta~\cite{wolf2408magnetism}.

Discretizing the continuum free energy on a lattice 
\begin{align}
    \int d^2r
\;\longrightarrow\;
A_{\rm cell}\sum_i ,\quad  \vec S(\vec r_i+\vec \delta_\eta)-\vec S(r_i)\sim( \vec \delta_\eta\cdot \vec \nabla) \vec S(\vec r_i).
\end{align}
where $A_{\rm cell}$ is the area of the lattice unit cell, $\vec r_i$ is the lattice position, and $\vec \delta_\eta
=
\delta_0
\left(
\cos\frac{2\pi\eta}{z},
\sin\frac{2\pi\eta}{z}
\right)$ is the bond between two neighboring lattice points. The coordination number $z$ depends on the lattice symmetry. Using the relation
\begin{align}
    \sum_{\eta=1}^z \left|( \vec \delta_\eta\cdot \vec \nabla) \vec S(\vec r_i)\right|^2=\frac{z\delta_0^2}{2}|\vec \nabla \vec S(\vec r_i)|^2,
\end{align}
we obtain
\begin{align}
    |\vec \nabla \vec S(\vec r)|^2\longrightarrow \frac{2}{z\delta_0^2}\sum_{\eta}|\vec S(\vec r_i+\vec \delta_\eta)-\vec S(r_i)|^2.
\end{align}

Using this, we construct the free-energy density form on a lattice
\begin{align}
    \delta F
    &=
    \frac{\rho_s A_{\rm cell}}{ z\delta_0^2S_0^2} \sum_i\sum_{\eta=1}^z \left|\vec S(\vec r_i+\vec \delta_\eta)-\vec S(\vec r_i)\right|^2\\
    &=-\frac{2\rho_s A_{\rm cell}}{ z\delta_0^2S_0^2} \sum_i\sum_{\eta=1}^z\vec S(\vec r_i+\vec \delta_\eta)\cdot \vec S(\vec r_i) +\rm const.
\end{align}
Dropping the constant and writing
$\vec S(\vec r_i)\equiv \vec S_i$, one obtains classical Heisenberg equation expressed in a symmetric form,
\begin{align}
    \delta F
    =
    -\frac{4\rho_s A_{\rm cell}}{ z\delta_0^2S_0^2}
    \sum_{\langle i,j\rangle}
    \vec S_i\cdot\vec S_j,
\end{align}
where $\sum_{\braket{i,j}}$ sums over nearest-neighbor bonds. Comparing with the effective lattice Hamiltonian \cref{eq:H_HM_main},
\begin{align}
    \hat H =-J_s \sum_{\braket{i,j}}\hat {\vec s}_i\cdot \hat {\vec s}_j+...,
\end{align}
we identify (taking $S_0=1$)
\begin{align}
    J_s=\frac{4\rho_s A_{\rm cell}}{ z\delta_0^2}.
\end{align}
In \cref{tab:J_s_rho_w}, we list the form of $J_s$ for different lattice geometry.  Throughout the main text, we use square lattice to construct $J_s$ and $J_\tau$ employed in the quantum Monte Carlo simulations. Varying the coarse-graining length within physically reasonable bounds changes the resulting ordering temperatures by less than $20\%$, indicating that the results are controlled primarily by the long-wavelength stiffness rather than lattice-scale details.
\begin{table}[h]
\centering
\caption{Exchange coupling for different lattice geometries. In the quantum Monte Carlo simulations we use a square lattice, for which $J_s=\rho_s$. The corresponding couplings are $J_s^{\rm QM}=0.10~\mathrm{meV}$, $J_s^{\rm HM}=0.30~\mathrm{meV}$, and $J_\tau^{\rm QM}=0.15~\mathrm{meV}$.
}
\label{tab:J_s_rho_w}
\begin{tabular}{@{}lccc@{}}
\hline
\hline
\textbf{Lattice} & \textbf{Coordination number} $z$ & $A_{\rm cell}/\delta_0^2$ & $J_s/\rho_s$ \\
\hline
Square & 4 & 1 & 1\\
Triangular & 6 & $\frac{\sqrt{3}}{2}$ & $\frac{1}{\sqrt{3}}$\\
Honeycomb & 3 & $\frac{3\sqrt{3}}{4}$ & $\sqrt{3}$\\
Kagome & 4 & $\frac{2\sqrt{3}}{3}$ & $\frac{2}{\sqrt{3}}$\\
\hline
\hline
\end{tabular}
\end{table}

The valley stiffness $\rho_\tau$ is extracted in complete analogy by imposing a long-wavelength valley-pseudospin spiral. A small symmetry-breaking field $\hat{\mathcal H}_{\rm c}=b\hat\tau_x$ is introduced to select the direction of valley coherence and stabilize the self-consistent solution.
\cref{fig:spin_valley_stiffness}(i) shows the change in energy with spiral wave vector $|\vec q|$ in the quarter-metal phase. A quadratic fit gives $\rho_\tau=0.15~\mathrm{meV}$. For the square-lattice mapping used in the quantum Monte Carlo simulations, $J_\tau=\rho_\tau=0.15~\mathrm{meV}$.

\section{Half-metal canting criterion and dependence on intervalley coupling}
\label{appendix:hm_g_dependence}

The qualitative dependence of the half-metal state on the intervalley coupling $g_\perp$ can be understood from a zero-temperature minimization. At $h_x=0$, we consider a uniform valley-canted configuration
\begin{equation}
\vec s_K=(\sin\theta,0,\cos\theta),\qquad \vec s_{\bar K}=(\sin\theta,0,-\cos\theta).
\label{eq:app_canting_ansatz}
\end{equation}
For a spatially uniform state, the intravalley spins remain parallel, so the $J_s$ term contributes only a $\theta$-independent constant.  The angle-dependent energy per site is therefore
\begin{equation}
E(\theta)=g_\perp\cos(2\theta)-\lambda\cos\theta+\mathrm{const.}
\label{eq:app_canting_energy}
\end{equation}
Minimization gives
\begin{equation}
\frac{\partial E}{\partial\theta}=\sin\theta\left(\lambda-4g_\perp\cos\theta\right)=0.
\label{eq:app_canting_condition}
\end{equation}
The collinear stationary solutions satisfy $\sin\theta=0$. For $\lambda>0$, the relevant collinear minimum is $\theta=0$. A nontrivial canted solution satisfies
\begin{equation}
\cos\theta=\frac{\lambda}{4g_\perp},
\label{eq:app_canting_angle}
\end{equation}
and has a finite in-plane component only when
\begin{equation}
\lambda<4g_\perp.
\label{eq:app_canting_boundary}
\end{equation}
Thus $\lambda=4g_\perp$ marks the zero-temperature boundary where the canted solution merges with the collinear state. For $\lambda>4g_\perp$, the spins remain polarized along opposite $z$ directions in the two valleys at $h_x=0$. For $\lambda<4g_\perp$, the ground state develops a finite common in-plane component whose azimuthal direction provides a continuous $\mathrm{U}(1)$ degree of freedom. This canted regime therefore allows a zero-field Berezinskii--Kosterlitz--Thouless transition.

\begin{figure*}[t]
    \centering
    \includegraphics[width=\linewidth]
    {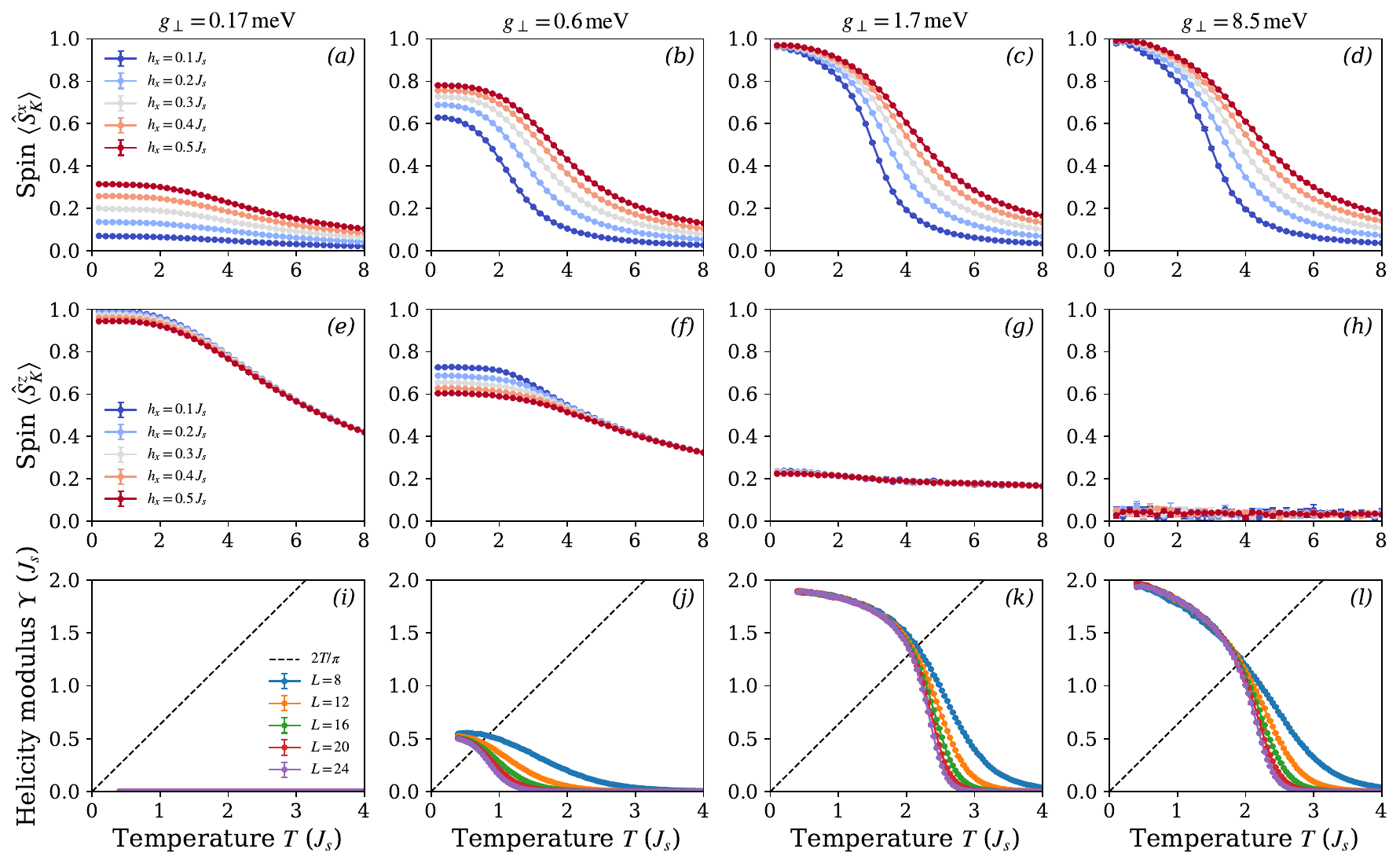}
   \caption{%
Dependence of the half-metal spin response and helicity modulus on the intervalley coupling $g_\perp$ at fixed $J_s=0.30~\mathrm{meV}$ and $\lambda=1.50~\mathrm{meV}$. The four columns correspond to $g_\perp=0.17,0.60,1.70,$ and $8.50~\mathrm{meV}$. (a--d) In-plane polarization $\langle \hat S_K^x\rangle$ and (e--h) out-of-plane polarization $\langle \hat S_K^z\rangle$ on an $L=16$ lattice for $h_x/J_s=0.1$--$0.5$. (i--l) Zero-field helicity modulus $\Upsilon/J_s$ for $L=8,12,16,20,$ and $24$. The dashed line denotes the universal Berezinskii--Kosterlitz--Thouless condition $\Upsilon=2T/\pi$.
}
    \label{fig:app_g_sweep}
\end{figure*}

For $\lambda=1.50~\mathrm{meV}$, the zero-temperature canting boundary is
\begin{equation}
g_\perp^c=\frac{\lambda}{4}=0.375~\mathrm{meV}.
\end{equation}
\Cref{fig:app_g_sweep} shows the corresponding Monte Carlo results for $g_\perp=0.17,0.60,1.70,$ and $8.50~\mathrm{meV}$.

The $g_\perp=0.17~\mathrm{meV}$ column provides a direct illustration of the canting criterion. A finite transverse field induces a nonzero $\langle \hat S_K^x\rangle$, but the zero-field helicity modulus remains essentially zero because $\lambda=1.50~\mathrm{meV}>4g_\perp=0.68~\mathrm{meV}$. In contrast, the other three values lie on the canted side of the boundary and exhibit a finite low-temperature helicity modulus together with a crossing of the universal-jump line. The value $g_\perp=0.60~\mathrm{meV}$ used in the main-text figures is therefore a representative point in the canted regime, relatively close to the onset at $g_\perp^c=0.375~\mathrm{meV}$.

\section{Finite-size scaling and universality}
\label{appendix:fss}

This appendix collects the finite-size evidence for the two transition assignments used in \cref{sec:results}: a 2D-Ising transition in the quarter metal and a Berezinskii--Kosterlitz--Thouless transition in the proximitized half-metal.

\subsection{Quarter-metal Binder crossings}

For the quarter-metal transition we use the spatially averaged spin polarization $\hat S^z$ as the Ising order parameter and compute the Binder ratio
\begin{equation}
Q_L(T)=\frac{\langle(\hat S^z)^2\rangle_L^2}{\langle(\hat S^z)^4\rangle_L}.
\label{eq:binder}
\end{equation}
In the ordered and disordered phases, $Q_L$ approaches $1$ and $1/3$, respectively, while curves for different $L$ cross near $T_c$. \Cref{fig:qm_binder} (main text) shows results for $L=8,12,16,24,$ and $40$ at $J_s=0.10~\mathrm{meV}$, $J_\tau=0.15~\mathrm{meV}$, and $\lambda=0.05~\mathrm{meV}$. The larger-system crossings concentrate at $T_c\approx0.72$--$0.75~\mathrm{K}$, consistent with the $T_c\approx0.7~\mathrm{K}$ scale quoted in the main text.

\begin{figure*}
  \centering
  \includegraphics[width=0.72\linewidth]{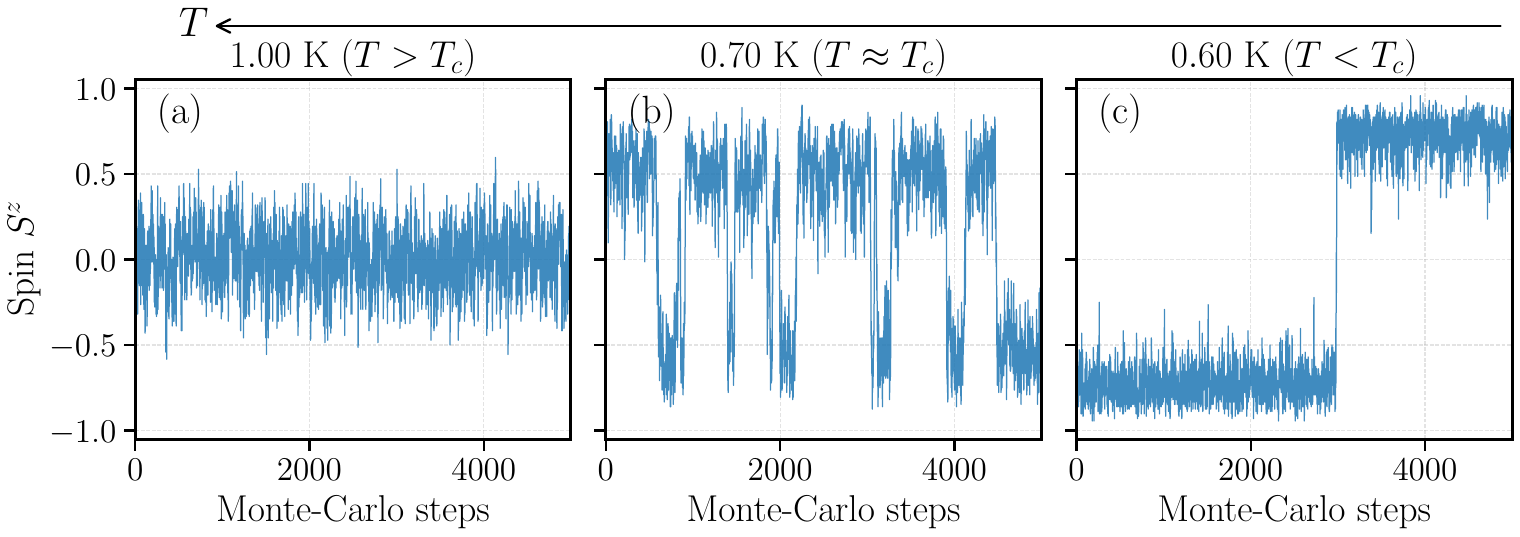}
  \caption{Monte Carlo histories at $L=12$ of the signed spin polarization $S^z$ per configuration (saturation 1): (a) $1.00$~K above $T_c$, (b) $0.70$~K near $T_c$, and (c) $0.60$~K below $T_c$.}
  \label{fig:qm_mc_histories}
\end{figure*}

\paragraph{Z$_2$ tunneling in raw MC trajectories.}
\Cref{fig:qm_mc_histories} shows histories of the signed spin polarization at $L=12$ for three representative temperatures; each trace records the value $S^z$ of $\hat S^z$ in the sampled configuration, sweep by sweep. Above $T_c$ the polarization fluctuates around zero, near $T_c$ the system tunnels between the two symmetry-related sectors, and below $T_c$ it remains trapped in one sector for long intervals. The local spin-valley locking $\hat C^{zz}$ is invariant under these joint $\mathbb Z_2$ flips, so its thermal average is insensitive to this global tunneling; as discussed in the main text, it measures the local locking between the spin and valley polarizations rather than long-range order by itself.

\subsection{Half-metal Berezinskii--Kosterlitz--Thouless extraction}

\begin{figure}[t]
  \centering
  \includegraphics[width=\linewidth]{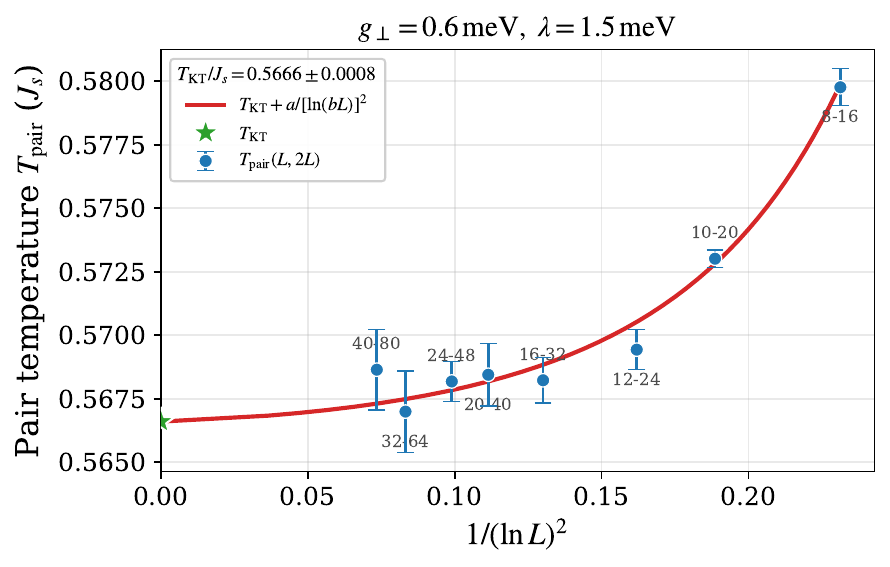}
\caption{%
Two-size finite-size extrapolation of the Berezinskii--Kosterlitz--Thouless transition temperature for $g_\perp=0.60~\mathrm{meV}$ and $\lambda=1.50~\mathrm{meV}$. The points show the pairwise estimates $T_{\rm pair}(L,2L)\equiv T^*(L)$ obtained from the two-size helicity analysis. The solid curve is a fit to $T^*(L)=T_{\rm KT}+a/[\ln(bL)]^2$. The thermodynamic-limit extrapolation gives $T_{\rm KT}/J_s=0.5666\pm0.0008$, corresponding to $T_{\rm KT}=1.973\pm0.003~\mathrm{K}$ for $J_s=0.30~\mathrm{meV}$.
}
  \label{fig:hm_helicity_fss}
\end{figure}

For the proximitized half-metal, we determine the Berezinskii--Kosterlitz--Thouless transition using the two-size finite-size scaling method of Ref.~\cite{hsieh2013finite}. In the thermodynamic limit, the physical spin stiffness satisfies the Nelson--Kosterlitz relation $\rho_s(T_{\rm KT})=2T_{\rm KT}/\pi$~\cite{nelson1977universal}. For a pair of system sizes $(L_1,L_2)$, the finite-size transition temperature $T^*(L_1,L_2)$ is defined by
\begin{equation}
\frac{\Upsilon(T^*,L_1)}{F_1(L_1)}
=\frac{\Upsilon(T^*,L_2)}{F_1(L_2)}
=f\frac{2T^*}{\pi},
\label{eq:app_two_size}
\end{equation}
where
\begin{align}
F_1(L)&=1+\frac{g}{2\ln L+C},\\
g&=1.00202783,\qquad f=0.99982471.
\label{eq:app_F1}
\end{align}
Here $f$ is the stiffness-renormalization factor for a square torus with periodic boundary conditions. We use pairs $(L,2L)$ and determine a separate $C=C(L,2L)$ for each pair so that \cref{eq:app_two_size} is satisfied at a common temperature $T^*(L)\equiv T^*(L,2L)$.

The remaining finite-size corrections are extrapolated using
\begin{equation}
T^*(L)=T_{\rm KT}+\frac{a}{[\ln(bL)]^2},
\label{eq:app_two_size_extrap}
\end{equation}
where $T_{\rm KT}$, $a$, and $b$ are fit parameters. We denote the pairwise estimates shown in \cref{fig:hm_helicity_fss} by $T_{\rm pair}(L,2L)\equiv T^*(L)$.


\medskip


\begin{thebibliography}{33}%
\makeatletter
\providecommand \@ifxundefined [1]{%
 \@ifx{#1\undefined}
}%
\providecommand \@ifnum [1]{%
 \ifnum #1\expandafter \@firstoftwo
 \else \expandafter \@secondoftwo
 \fi
}%
\providecommand \@ifx [1]{%
 \ifx #1\expandafter \@firstoftwo
 \else \expandafter \@secondoftwo
 \fi
}%
\providecommand \natexlab [1]{#1}%
\providecommand \enquote  [1]{``#1''}%
\providecommand \bibnamefont  [1]{#1}%
\providecommand \bibfnamefont [1]{#1}%
\providecommand \citenamefont [1]{#1}%
\providecommand \href@noop [0]{\@secondoftwo}%
\providecommand \href [0]{\begingroup \@sanitize@url \@href}%
\providecommand \@href[1]{\@@startlink{#1}\@@href}%
\providecommand \@@href[1]{\endgroup#1\@@endlink}%
\providecommand \@sanitize@url [0]{\catcode `\\12\catcode `\$12\catcode
  `\&12\catcode `\#12\catcode `\^12\catcode `\_12\catcode `\%12\relax}%
\providecommand \@@startlink[1]{}%
\providecommand \@@endlink[0]{}%
\providecommand \url  [0]{\begingroup\@sanitize@url \@url }%
\providecommand \@url [1]{\endgroup\@href {#1}{\urlprefix }}%
\providecommand \urlprefix  [0]{URL }%
\providecommand \Eprint [0]{\href }%
\providecommand \doibase [0]{https://doi.org/}%
\providecommand \selectlanguage [0]{\@gobble}%
\providecommand \bibinfo  [0]{\@secondoftwo}%
\providecommand \bibfield  [0]{\@secondoftwo}%
\providecommand \translation [1]{[#1]}%
\providecommand \BibitemOpen [0]{}%
\providecommand \bibitemStop [0]{}%
\providecommand \bibitemNoStop [0]{.\EOS\space}%
\providecommand \EOS [0]{\spacefactor3000\relax}%
\providecommand \BibitemShut  [1]{\csname bibitem#1\endcsname}%
\let\auto@bib@innerbib\@empty
\bibitem [{\citenamefont {Zhou}\ \emph
  {et~al.}(2021{\natexlab{a}})\citenamefont {Zhou}, \citenamefont {Xie},
  \citenamefont {Ghazaryan}, \citenamefont {Holder}, \citenamefont {Ehrets},
  \citenamefont {Spanton}, \citenamefont {Taniguchi}, \citenamefont {Watanabe},
  \citenamefont {Berg}, \citenamefont {Serbyn},\ and\ \citenamefont
  {Young}}]{zhou_half_2021}%
  \BibitemOpen
  \bibfield  {author} {\bibinfo {author} {\bibfnamefont {H.}~\bibnamefont
  {Zhou}}, \bibinfo {author} {\bibfnamefont {T.}~\bibnamefont {Xie}}, \bibinfo
  {author} {\bibfnamefont {A.}~\bibnamefont {Ghazaryan}}, \bibinfo {author}
  {\bibfnamefont {T.}~\bibnamefont {Holder}}, \bibinfo {author} {\bibfnamefont
  {J.~R.}\ \bibnamefont {Ehrets}}, \bibinfo {author} {\bibfnamefont {E.~M.}\
  \bibnamefont {Spanton}}, \bibinfo {author} {\bibfnamefont {T.}~\bibnamefont
  {Taniguchi}}, \bibinfo {author} {\bibfnamefont {K.}~\bibnamefont {Watanabe}},
  \bibinfo {author} {\bibfnamefont {E.}~\bibnamefont {Berg}}, \bibinfo {author}
  {\bibfnamefont {M.}~\bibnamefont {Serbyn}},\ and\ \bibinfo {author}
  {\bibfnamefont {A.~F.}\ \bibnamefont {Young}},\ }\bibfield  {title} {\bibinfo
  {title} {Half and quarter metals in rhombohedral trilayer graphene},\ }\href
  {https://doi.org/10.1038/s41586-021-03938-w} {\bibfield  {journal} {\bibinfo
  {journal} {Nature}\ }\textbf {\bibinfo {volume} {598}},\ \bibinfo {pages}
  {429} (\bibinfo {year} {2021}{\natexlab{a}})},\ \Eprint
  {https://arxiv.org/abs/2104.00653} {arXiv:2104.00653} \BibitemShut {NoStop}%
\bibitem [{\citenamefont {Han}\ \emph {et~al.}(2023)\citenamefont {Han},
  \citenamefont {Lu}, \citenamefont {Scuri}, \citenamefont {Sung},
  \citenamefont {Wang}, \citenamefont {Han}, \citenamefont {Watanabe},
  \citenamefont {Taniguchi}, \citenamefont {Fu}, \citenamefont {Park},\ and\
  \citenamefont {Ju}}]{Han2023}%
  \BibitemOpen
  \bibfield  {author} {\bibinfo {author} {\bibfnamefont {T.}~\bibnamefont
  {Han}}, \bibinfo {author} {\bibfnamefont {Z.}~\bibnamefont {Lu}}, \bibinfo
  {author} {\bibfnamefont {G.}~\bibnamefont {Scuri}}, \bibinfo {author}
  {\bibfnamefont {J.}~\bibnamefont {Sung}}, \bibinfo {author} {\bibfnamefont
  {J.}~\bibnamefont {Wang}}, \bibinfo {author} {\bibfnamefont {T.}~\bibnamefont
  {Han}}, \bibinfo {author} {\bibfnamefont {K.}~\bibnamefont {Watanabe}},
  \bibinfo {author} {\bibfnamefont {T.}~\bibnamefont {Taniguchi}}, \bibinfo
  {author} {\bibfnamefont {L.}~\bibnamefont {Fu}}, \bibinfo {author}
  {\bibfnamefont {H.}~\bibnamefont {Park}},\ and\ \bibinfo {author}
  {\bibfnamefont {L.}~\bibnamefont {Ju}},\ }\bibfield  {title} {\bibinfo
  {title} {Orbital multiferroicity in pentalayer rhombohedral graphene},\
  }\href {https://doi.org/10.1038/s41586-023-06572-w} {\bibfield  {journal}
  {\bibinfo  {journal} {Nature}\ }\textbf {\bibinfo {volume} {623}},\ \bibinfo
  {pages} {41} (\bibinfo {year} {2023})}\BibitemShut {NoStop}%
\bibitem [{\citenamefont {Lu}\ \emph {et~al.}(2024)\citenamefont {Lu},
  \citenamefont {Han}, \citenamefont {Yao}, \citenamefont {Reddy},
  \citenamefont {Yang}, \citenamefont {Seo}, \citenamefont {Watanabe},
  \citenamefont {Taniguchi}, \citenamefont {Fu},\ and\ \citenamefont
  {Ju}}]{Lu2024}%
  \BibitemOpen
  \bibfield  {author} {\bibinfo {author} {\bibfnamefont {Z.}~\bibnamefont
  {Lu}}, \bibinfo {author} {\bibfnamefont {T.}~\bibnamefont {Han}}, \bibinfo
  {author} {\bibfnamefont {Y.}~\bibnamefont {Yao}}, \bibinfo {author}
  {\bibfnamefont {A.~P.}\ \bibnamefont {Reddy}}, \bibinfo {author}
  {\bibfnamefont {J.}~\bibnamefont {Yang}}, \bibinfo {author} {\bibfnamefont
  {J.}~\bibnamefont {Seo}}, \bibinfo {author} {\bibfnamefont {K.}~\bibnamefont
  {Watanabe}}, \bibinfo {author} {\bibfnamefont {T.}~\bibnamefont {Taniguchi}},
  \bibinfo {author} {\bibfnamefont {L.}~\bibnamefont {Fu}},\ and\ \bibinfo
  {author} {\bibfnamefont {L.}~\bibnamefont {Ju}},\ }\bibfield  {title}
  {\bibinfo {title} {Fractional quantum anomalous hall effect in multilayer
  graphene},\ }\href {https://doi.org/10.1038/s41586-023-07010-7} {\bibfield
  {journal} {\bibinfo  {journal} {Nature}\ }\textbf {\bibinfo {volume} {626}},\
  \bibinfo {pages} {759} (\bibinfo {year} {2024})}\BibitemShut {NoStop}%
\bibitem [{\citenamefont {Han}\ \emph {et~al.}(2025)\citenamefont {Han},
  \citenamefont {Lu}, \citenamefont {Hadjri}, \citenamefont {Shi},
  \citenamefont {Wu}, \citenamefont {Xu}, \citenamefont {Yao}, \citenamefont
  {Cotten}, \citenamefont {Sedeh}, \citenamefont {Weldeyesus}, \citenamefont
  {Yang}, \citenamefont {Seo}, \citenamefont {Ye}, \citenamefont {Zhou},
  \citenamefont {Liu}, \citenamefont {Shi}, \citenamefont {Hua}, \citenamefont
  {Watanabe}, \citenamefont {Taniguchi}, \citenamefont {Xiong}, \citenamefont
  {Zumb\"uhl}, \citenamefont {Fu},\ and\ \citenamefont {Ju}}]{Han2025}%
  \BibitemOpen
  \bibfield  {author} {\bibinfo {author} {\bibfnamefont {T.}~\bibnamefont
  {Han}}, \bibinfo {author} {\bibfnamefont {Z.}~\bibnamefont {Lu}}, \bibinfo
  {author} {\bibfnamefont {Z.}~\bibnamefont {Hadjri}}, \bibinfo {author}
  {\bibfnamefont {L.}~\bibnamefont {Shi}}, \bibinfo {author} {\bibfnamefont
  {Z.}~\bibnamefont {Wu}}, \bibinfo {author} {\bibfnamefont {W.}~\bibnamefont
  {Xu}}, \bibinfo {author} {\bibfnamefont {Y.}~\bibnamefont {Yao}}, \bibinfo
  {author} {\bibfnamefont {A.~A.}\ \bibnamefont {Cotten}}, \bibinfo {author}
  {\bibfnamefont {O.~S.}\ \bibnamefont {Sedeh}}, \bibinfo {author}
  {\bibfnamefont {H.}~\bibnamefont {Weldeyesus}}, \bibinfo {author}
  {\bibfnamefont {J.}~\bibnamefont {Yang}}, \bibinfo {author} {\bibfnamefont
  {J.}~\bibnamefont {Seo}}, \bibinfo {author} {\bibfnamefont {S.}~\bibnamefont
  {Ye}}, \bibinfo {author} {\bibfnamefont {M.}~\bibnamefont {Zhou}}, \bibinfo
  {author} {\bibfnamefont {H.}~\bibnamefont {Liu}}, \bibinfo {author}
  {\bibfnamefont {G.}~\bibnamefont {Shi}}, \bibinfo {author} {\bibfnamefont
  {Z.}~\bibnamefont {Hua}}, \bibinfo {author} {\bibfnamefont {K.}~\bibnamefont
  {Watanabe}}, \bibinfo {author} {\bibfnamefont {T.}~\bibnamefont {Taniguchi}},
  \bibinfo {author} {\bibfnamefont {P.}~\bibnamefont {Xiong}}, \bibinfo
  {author} {\bibfnamefont {D.~M.}\ \bibnamefont {Zumb\"uhl}}, \bibinfo {author}
  {\bibfnamefont {L.}~\bibnamefont {Fu}},\ and\ \bibinfo {author}
  {\bibfnamefont {L.}~\bibnamefont {Ju}},\ }\bibfield  {title} {\bibinfo
  {title} {Signatures of chiral superconductivity in rhombohedral graphene},\
  }\href {https://doi.org/10.1038/s41586-025-09169-7} {\bibfield  {journal}
  {\bibinfo  {journal} {Nature}\ }\textbf {\bibinfo {volume} {643}},\ \bibinfo
  {pages} {654} (\bibinfo {year} {2025})}\BibitemShut {NoStop}%
\bibitem [{\citenamefont {Geier}\ \emph {et~al.}(2026)\citenamefont {Geier},
  \citenamefont {Davydova},\ and\ \citenamefont {Fu}}]{Geier2026}%
  \BibitemOpen
  \bibfield  {author} {\bibinfo {author} {\bibfnamefont {M.}~\bibnamefont
  {Geier}}, \bibinfo {author} {\bibfnamefont {M.}~\bibnamefont {Davydova}},\
  and\ \bibinfo {author} {\bibfnamefont {L.}~\bibnamefont {Fu}},\ }\bibfield
  {title} {\bibinfo {title} {Chiral and topological superconductivity in
  isospin polarized multilayer graphene},\ }\href
  {https://doi.org/10.1038/s41467-025-66902-6} {\bibfield  {journal} {\bibinfo
  {journal} {Nat. Commun.}\ }\textbf {\bibinfo {volume} {17}},\ \bibinfo
  {pages} {232} (\bibinfo {year} {2026})}\BibitemShut {NoStop}%
\bibitem [{\citenamefont {Ghazaryan}\ \emph {et~al.}(2023)\citenamefont
  {Ghazaryan}, \citenamefont {Holder}, \citenamefont {Berg},\ and\
  \citenamefont {Serbyn}}]{Ghazaryan2023}%
  \BibitemOpen
  \bibfield  {author} {\bibinfo {author} {\bibfnamefont {A.}~\bibnamefont
  {Ghazaryan}}, \bibinfo {author} {\bibfnamefont {T.}~\bibnamefont {Holder}},
  \bibinfo {author} {\bibfnamefont {E.}~\bibnamefont {Berg}},\ and\ \bibinfo
  {author} {\bibfnamefont {M.}~\bibnamefont {Serbyn}},\ }\bibfield  {title}
  {\bibinfo {title} {Multilayer graphenes as a platform for interaction-driven
  physics and topological superconductivity},\ }\href
  {https://doi.org/10.1103/PhysRevB.107.104502} {\bibfield  {journal} {\bibinfo
   {journal} {Phys. Rev. B}\ }\textbf {\bibinfo {volume} {107}},\ \bibinfo
  {pages} {104502} (\bibinfo {year} {2023})}\BibitemShut {NoStop}%
\bibitem [{\citenamefont {Slizovskiy}\ \emph {et~al.}(2019)\citenamefont
  {Slizovskiy}, \citenamefont {McCann}, \citenamefont {Koshino},\ and\
  \citenamefont {Fal'ko}}]{Slizovskiy2019}%
  \BibitemOpen
  \bibfield  {author} {\bibinfo {author} {\bibfnamefont {S.}~\bibnamefont
  {Slizovskiy}}, \bibinfo {author} {\bibfnamefont {E.}~\bibnamefont {McCann}},
  \bibinfo {author} {\bibfnamefont {M.}~\bibnamefont {Koshino}},\ and\ \bibinfo
  {author} {\bibfnamefont {V.~I.}\ \bibnamefont {Fal'ko}},\ }\bibfield  {title}
  {\bibinfo {title} {Films of rhombohedral graphite as two-dimensional
  topological semimetals},\ }\href {https://doi.org/10.1038/s42005-019-0268-8}
  {\bibfield  {journal} {\bibinfo  {journal} {Commun. Phys.}\ }\textbf
  {\bibinfo {volume} {2}},\ \bibinfo {pages} {164} (\bibinfo {year}
  {2019})}\BibitemShut {NoStop}%
\bibitem [{\citenamefont {Zhou}\ \emph
  {et~al.}(2021{\natexlab{b}})\citenamefont {Zhou}, \citenamefont {Xie},
  \citenamefont {Taniguchi}, \citenamefont {Watanabe},\ and\ \citenamefont
  {Young}}]{Zhou_superconductivity_2021}%
  \BibitemOpen
  \bibfield  {author} {\bibinfo {author} {\bibfnamefont {H.}~\bibnamefont
  {Zhou}}, \bibinfo {author} {\bibfnamefont {T.}~\bibnamefont {Xie}}, \bibinfo
  {author} {\bibfnamefont {T.}~\bibnamefont {Taniguchi}}, \bibinfo {author}
  {\bibfnamefont {K.}~\bibnamefont {Watanabe}},\ and\ \bibinfo {author}
  {\bibfnamefont {A.~F.}\ \bibnamefont {Young}},\ }\bibfield  {title} {\bibinfo
  {title} {Superconductivity in rhombohedral trilayer graphene},\ }\href
  {https://doi.org/10.1038/s41586-021-03926-0} {\bibfield  {journal} {\bibinfo
  {journal} {Nature}\ }\textbf {\bibinfo {volume} {598}},\ \bibinfo {pages}
  {434} (\bibinfo {year} {2021}{\natexlab{b}})}\BibitemShut {NoStop}%
\bibitem [{\citenamefont {Han}\ \emph {et~al.}(2024)\citenamefont {Han},
  \citenamefont {Lu}, \citenamefont {Yao}, \citenamefont {Yang}, \citenamefont
  {Seo}, \citenamefont {Yoon}, \citenamefont {Watanabe}, \citenamefont
  {Taniguchi}, \citenamefont {Fu}, \citenamefont {Zhang},\ and\ \citenamefont
  {Ju}}]{Han2024}%
  \BibitemOpen
  \bibfield  {author} {\bibinfo {author} {\bibfnamefont {T.}~\bibnamefont
  {Han}}, \bibinfo {author} {\bibfnamefont {Z.}~\bibnamefont {Lu}}, \bibinfo
  {author} {\bibfnamefont {Y.}~\bibnamefont {Yao}}, \bibinfo {author}
  {\bibfnamefont {J.}~\bibnamefont {Yang}}, \bibinfo {author} {\bibfnamefont
  {J.}~\bibnamefont {Seo}}, \bibinfo {author} {\bibfnamefont {C.}~\bibnamefont
  {Yoon}}, \bibinfo {author} {\bibfnamefont {K.}~\bibnamefont {Watanabe}},
  \bibinfo {author} {\bibfnamefont {T.}~\bibnamefont {Taniguchi}}, \bibinfo
  {author} {\bibfnamefont {L.}~\bibnamefont {Fu}}, \bibinfo {author}
  {\bibfnamefont {F.}~\bibnamefont {Zhang}},\ and\ \bibinfo {author}
  {\bibfnamefont {L.}~\bibnamefont {Ju}},\ }\bibfield  {title} {\bibinfo
  {title} {Large quantum anomalous hall effect in spin-orbit proximitized
  rhombohedral graphene},\ }\href {https://doi.org/10.1126/science.adk9749}
  {\bibfield  {journal} {\bibinfo  {journal} {Science}\ }\textbf {\bibinfo
  {volume} {384}},\ \bibinfo {pages} {647} (\bibinfo {year}
  {2024})}\BibitemShut {NoStop}%
\bibitem [{\citenamefont {Yin}\ \emph {et~al.}(2022)\citenamefont {Yin},
  \citenamefont {Tan}, \citenamefont {Barcons-Ruiz}, \citenamefont {Torre},
  \citenamefont {Watanabe}, \citenamefont {Taniguchi}, \citenamefont {Song},
  \citenamefont {Hone},\ and\ \citenamefont {Koppens}}]{Yin2022}%
  \BibitemOpen
  \bibfield  {author} {\bibinfo {author} {\bibfnamefont {J.}~\bibnamefont
  {Yin}}, \bibinfo {author} {\bibfnamefont {C.}~\bibnamefont {Tan}}, \bibinfo
  {author} {\bibfnamefont {D.}~\bibnamefont {Barcons-Ruiz}}, \bibinfo {author}
  {\bibfnamefont {I.}~\bibnamefont {Torre}}, \bibinfo {author} {\bibfnamefont
  {K.}~\bibnamefont {Watanabe}}, \bibinfo {author} {\bibfnamefont
  {T.}~\bibnamefont {Taniguchi}}, \bibinfo {author} {\bibfnamefont {J.~C.~W.}\
  \bibnamefont {Song}}, \bibinfo {author} {\bibfnamefont {J.}~\bibnamefont
  {Hone}},\ and\ \bibinfo {author} {\bibfnamefont {F.~H.~L.}\ \bibnamefont
  {Koppens}},\ }\bibfield  {title} {\bibinfo {title} {Tunable and giant
  valley-selective hall effect in gapped bilayer graphene},\ }\href
  {https://doi.org/10.1126/science.abl4266} {\bibfield  {journal} {\bibinfo
  {journal} {Science}\ }\textbf {\bibinfo {volume} {375}},\ \bibinfo {pages}
  {1398} (\bibinfo {year} {2022})}\BibitemShut {NoStop}%
\bibitem [{\citenamefont {Huang}\ \emph {et~al.}(2023)\citenamefont {Huang},
  \citenamefont {Wolf}, \citenamefont {Qin}, \citenamefont {Wei}, \citenamefont
  {Blinov},\ and\ \citenamefont {MacDonald}}]{huang2023spin}%
  \BibitemOpen
  \bibfield  {author} {\bibinfo {author} {\bibfnamefont {C.}~\bibnamefont
  {Huang}}, \bibinfo {author} {\bibfnamefont {T.~M.}\ \bibnamefont {Wolf}},
  \bibinfo {author} {\bibfnamefont {W.}~\bibnamefont {Qin}}, \bibinfo {author}
  {\bibfnamefont {N.}~\bibnamefont {Wei}}, \bibinfo {author} {\bibfnamefont
  {I.~V.}\ \bibnamefont {Blinov}},\ and\ \bibinfo {author} {\bibfnamefont
  {A.~H.}\ \bibnamefont {MacDonald}},\ }\bibfield  {title} {\bibinfo {title}
  {Spin and orbital metallic magnetism in rhombohedral trilayer graphene},\
  }\href@noop {} {\bibfield  {journal} {\bibinfo  {journal} {Phys. Rev. B}\
  }\textbf {\bibinfo {volume} {107}},\ \bibinfo {pages} {L121405} (\bibinfo
  {year} {2023})}\BibitemShut {NoStop}%
\bibitem [{\citenamefont {Holleis}\ \emph {et~al.}(2025)\citenamefont
  {Holleis}, \citenamefont {Xie}, \citenamefont {Xu}, \citenamefont {Zhou},
  \citenamefont {Patterson}, \citenamefont {Panigrahi}, \citenamefont
  {Taniguchi}, \citenamefont {Watanabe}, \citenamefont {Levitov}, \citenamefont
  {Jin}, \citenamefont {Berg},\ and\ \citenamefont {Young}}]{Holleis_2025}%
  \BibitemOpen
  \bibfield  {author} {\bibinfo {author} {\bibfnamefont {L.}~\bibnamefont
  {Holleis}}, \bibinfo {author} {\bibfnamefont {T.}~\bibnamefont {Xie}},
  \bibinfo {author} {\bibfnamefont {S.}~\bibnamefont {Xu}}, \bibinfo {author}
  {\bibfnamefont {H.}~\bibnamefont {Zhou}}, \bibinfo {author} {\bibfnamefont
  {C.~L.}\ \bibnamefont {Patterson}}, \bibinfo {author} {\bibfnamefont
  {A.}~\bibnamefont {Panigrahi}}, \bibinfo {author} {\bibfnamefont
  {T.}~\bibnamefont {Taniguchi}}, \bibinfo {author} {\bibfnamefont
  {K.}~\bibnamefont {Watanabe}}, \bibinfo {author} {\bibfnamefont {L.~S.}\
  \bibnamefont {Levitov}}, \bibinfo {author} {\bibfnamefont {C.}~\bibnamefont
  {Jin}}, \bibinfo {author} {\bibfnamefont {E.}~\bibnamefont {Berg}},\ and\
  \bibinfo {author} {\bibfnamefont {A.~F.}\ \bibnamefont {Young}},\ }\bibfield
  {title} {\bibinfo {title} {Fluctuating magnetism and {Pomeranchuk} effect in
  multilayer graphene},\ }\href {https://doi.org/10.1038/s41586-025-08725-5}
  {\bibfield  {journal} {\bibinfo  {journal} {Nature}\ }\textbf {\bibinfo
  {volume} {640}},\ \bibinfo {pages} {355} (\bibinfo {year}
  {2025})}\BibitemShut {NoStop}%
\bibitem [{\citenamefont {Doniach}\ and\ \citenamefont
  {Engelsberg}(1966)}]{doniach1966low}%
  \BibitemOpen
  \bibfield  {author} {\bibinfo {author} {\bibfnamefont {S.}~\bibnamefont
  {Doniach}}\ and\ \bibinfo {author} {\bibfnamefont {S.}~\bibnamefont
  {Engelsberg}},\ }\bibfield  {title} {\bibinfo {title} {Low-temperature
  properties of nearly ferromagnetic {Fermi} liquids},\ }\href@noop {}
  {\bibfield  {journal} {\bibinfo  {journal} {Phys. Rev. Lett.}\ }\textbf
  {\bibinfo {volume} {17}},\ \bibinfo {pages} {750} (\bibinfo {year}
  {1966})}\BibitemShut {NoStop}%
\bibitem [{\citenamefont {Vollhardt}(1984)}]{vollhardt1984normal}%
  \BibitemOpen
  \bibfield  {author} {\bibinfo {author} {\bibfnamefont {D.}~\bibnamefont
  {Vollhardt}},\ }\bibfield  {title} {\bibinfo {title} {Normal
  $^{3}\mathrm{He}$: an almost localized fermi liquid},\ }\href
  {https://doi.org/10.1103/RevModPhys.56.99} {\bibfield  {journal} {\bibinfo
  {journal} {Rev. Mod. Phys.}\ }\textbf {\bibinfo {volume} {56}},\ \bibinfo
  {pages} {99} (\bibinfo {year} {1984})}\BibitemShut {NoStop}%
\bibitem [{\citenamefont {Georges}\ and\ \citenamefont
  {Laloux}(1997)}]{georges1997normal}%
  \BibitemOpen
  \bibfield  {author} {\bibinfo {author} {\bibfnamefont {A.}~\bibnamefont
  {Georges}}\ and\ \bibinfo {author} {\bibfnamefont {L.}~\bibnamefont
  {Laloux}},\ }\bibfield  {title} {\bibinfo {title} {Normal helium 3: A
  mott–stoner liquid},\ }\href {https://doi.org/10.1142/s0217984997001122}
  {\bibfield  {journal} {\bibinfo  {journal} {Mod. Phys. Lett. B}\ }\textbf
  {\bibinfo {volume} {11}},\ \bibinfo {pages} {913} (\bibinfo {year}
  {1997})}\BibitemShut {NoStop}%
\bibitem [{\citenamefont {Sun}\ \emph {et~al.}(2026)\citenamefont {Sun},
  \citenamefont {Deng}, \citenamefont {Xie}, \citenamefont {Ge}, \citenamefont
  {Li}, \citenamefont {Taniguchi}, \citenamefont {Watanabe},\ and\
  \citenamefont {Liu}}]{sun2026real}%
  \BibitemOpen
  \bibfield  {author} {\bibinfo {author} {\bibfnamefont {Y.}~\bibnamefont
  {Sun}}, \bibinfo {author} {\bibfnamefont {J.}~\bibnamefont {Deng}}, \bibinfo
  {author} {\bibfnamefont {J.}~\bibnamefont {Xie}}, \bibinfo {author}
  {\bibfnamefont {D.}~\bibnamefont {Ge}}, \bibinfo {author} {\bibfnamefont
  {H.}~\bibnamefont {Li}}, \bibinfo {author} {\bibfnamefont {T.}~\bibnamefont
  {Taniguchi}}, \bibinfo {author} {\bibfnamefont {K.}~\bibnamefont
  {Watanabe}},\ and\ \bibinfo {author} {\bibfnamefont {X.}~\bibnamefont
  {Liu}},\ }\href@noop {} {\bibinfo {title} {Real-space visualization of
  emergent electron crystals in rhombohedral graphene}} (\bibinfo {year}
  {2026}),\ \Eprint {https://arxiv.org/abs/2609.12057} {arXiv:2609.12057
  [cond-mat.mes-hall]} \BibitemShut {NoStop}%
\bibitem [{\citenamefont {Wolf}\ \emph {et~al.}(2024)\citenamefont {Wolf},
  \citenamefont {Wei}, \citenamefont {Zhou},\ and\ \citenamefont
  {Huang}}]{wolf2408magnetism}%
  \BibitemOpen
  \bibfield  {author} {\bibinfo {author} {\bibfnamefont {T.}~\bibnamefont
  {Wolf}}, \bibinfo {author} {\bibfnamefont {N.}~\bibnamefont {Wei}}, \bibinfo
  {author} {\bibfnamefont {H.}~\bibnamefont {Zhou}},\ and\ \bibinfo {author}
  {\bibfnamefont {C.}~\bibnamefont {Huang}},\ }\href
  {https://arxiv.org/abs/2408.15884} {\bibinfo {title} {Magnetism in the dilute
  electron gas of rhombohedral multilayer graphene}} (\bibinfo {year} {2024}),\
  \Eprint {https://arxiv.org/abs/2408.15884} {arXiv:2408.15884
  [cond-mat.str-el]} \BibitemShut {NoStop}%
\bibitem [{\citenamefont {Moriya}(1979)}]{moriya1979recent}%
  \BibitemOpen
  \bibfield  {author} {\bibinfo {author} {\bibfnamefont {T.}~\bibnamefont
  {Moriya}},\ }\bibfield  {title} {\bibinfo {title} {Recent progress in the
  theory of itinerant electron magnetism},\ }\href@noop {} {\bibfield
  {journal} {\bibinfo  {journal} {J. Magn. Magn. Mater.}\ }\textbf {\bibinfo
  {volume} {14}},\ \bibinfo {pages} {1} (\bibinfo {year} {1979})}\BibitemShut
  {NoStop}%
\bibitem [{\citenamefont {Patterson}\ \emph {et~al.}(2025)\citenamefont
  {Patterson}, \citenamefont {Sheekey}, \citenamefont {Arp}, \citenamefont
  {Holleis}, \citenamefont {Koh}, \citenamefont {Choi}, \citenamefont {Xie},
  \citenamefont {Xu}, \citenamefont {Guo}, \citenamefont {Stoyanov} \emph
  {et~al.}}]{patterson2025superconductivity}%
  \BibitemOpen
  \bibfield  {author} {\bibinfo {author} {\bibfnamefont {C.~L.}\ \bibnamefont
  {Patterson}}, \bibinfo {author} {\bibfnamefont {O.~I.}\ \bibnamefont
  {Sheekey}}, \bibinfo {author} {\bibfnamefont {T.~B.}\ \bibnamefont {Arp}},
  \bibinfo {author} {\bibfnamefont {L.~F.}\ \bibnamefont {Holleis}}, \bibinfo
  {author} {\bibfnamefont {J.~M.}\ \bibnamefont {Koh}}, \bibinfo {author}
  {\bibfnamefont {Y.}~\bibnamefont {Choi}}, \bibinfo {author} {\bibfnamefont
  {T.}~\bibnamefont {Xie}}, \bibinfo {author} {\bibfnamefont {S.}~\bibnamefont
  {Xu}}, \bibinfo {author} {\bibfnamefont {Y.}~\bibnamefont {Guo}}, \bibinfo
  {author} {\bibfnamefont {H.}~\bibnamefont {Stoyanov}}, \emph {et~al.},\
  }\bibfield  {title} {\bibinfo {title} {Superconductivity and spin canting in
  spin--orbit-coupled trilayer graphene},\ }\href@noop {} {\bibfield  {journal}
  {\bibinfo  {journal} {Nature}\ }\textbf {\bibinfo {volume} {641}},\ \bibinfo
  {pages} {632} (\bibinfo {year} {2025})}\BibitemShut {NoStop}%
\bibitem [{\citenamefont {Dong}\ \emph {et~al.}(2026)\citenamefont {Dong},
  \citenamefont {Lantagne-Hurtubise},\ and\ \citenamefont
  {Alicea}}]{dong2026superconductivity}%
  \BibitemOpen
  \bibfield  {author} {\bibinfo {author} {\bibfnamefont {Z.}~\bibnamefont
  {Dong}}, \bibinfo {author} {\bibfnamefont {E.}~\bibnamefont
  {Lantagne-Hurtubise}},\ and\ \bibinfo {author} {\bibfnamefont
  {J.}~\bibnamefont {Alicea}},\ }\bibfield  {title} {\bibinfo {title}
  {Superconductivity from spin-canting fluctuations in rhombohedral graphene},\
  }\href {https://doi.org/10.1103/2ts4-cxhs} {\bibfield  {journal} {\bibinfo
  {journal} {Phys. Rev. X}\ }\textbf {\bibinfo {volume} {16}},\ \bibinfo
  {pages} {011055} (\bibinfo {year} {2026})}\BibitemShut {NoStop}%
\bibitem [{\citenamefont {Henelius}\ \emph {et~al.}(2000)\citenamefont
  {Henelius}, \citenamefont {Sandvik}, \citenamefont {Timm},\ and\
  \citenamefont {Girvin}}]{henelius2000monte}%
  \BibitemOpen
  \bibfield  {author} {\bibinfo {author} {\bibfnamefont {P.}~\bibnamefont
  {Henelius}}, \bibinfo {author} {\bibfnamefont {A.~W.}\ \bibnamefont
  {Sandvik}}, \bibinfo {author} {\bibfnamefont {C.}~\bibnamefont {Timm}},\ and\
  \bibinfo {author} {\bibfnamefont {S.}~\bibnamefont {Girvin}},\ }\bibfield
  {title} {\bibinfo {title} {{Monte Carlo} study of a two-dimensional quantum
  ferromagnet},\ }\href@noop {} {\bibfield  {journal} {\bibinfo  {journal}
  {Phys. Rev. B}\ }\textbf {\bibinfo {volume} {61}},\ \bibinfo {pages} {364}
  (\bibinfo {year} {2000})}\BibitemShut {NoStop}%
\bibitem [{\citenamefont {Henelius}\ \emph {et~al.}(2002)\citenamefont
  {Henelius}, \citenamefont {Fr{\"o}brich}, \citenamefont {Kuntz},
  \citenamefont {Timm},\ and\ \citenamefont {Jensen}}]{henelius2002quantum}%
  \BibitemOpen
  \bibfield  {author} {\bibinfo {author} {\bibfnamefont {P.}~\bibnamefont
  {Henelius}}, \bibinfo {author} {\bibfnamefont {P.}~\bibnamefont
  {Fr{\"o}brich}}, \bibinfo {author} {\bibfnamefont {P.}~\bibnamefont {Kuntz}},
  \bibinfo {author} {\bibfnamefont {C.}~\bibnamefont {Timm}},\ and\ \bibinfo
  {author} {\bibfnamefont {P.}~\bibnamefont {Jensen}},\ }\bibfield  {title}
  {\bibinfo {title} {Quantum {Monte Carlo} simulation of thin magnetic films},\
  }\href@noop {} {\bibfield  {journal} {\bibinfo  {journal} {Phys. Rev. B}\
  }\textbf {\bibinfo {volume} {66}},\ \bibinfo {pages} {094407} (\bibinfo
  {year} {2002})}\BibitemShut {NoStop}%
\bibitem [{\citenamefont {Herring}(1966)}]{herring1966exchange}%
  \BibitemOpen
  \bibfield  {author} {\bibinfo {author} {\bibfnamefont {C.}~\bibnamefont
  {Herring}},\ }\bibfield  {title} {\bibinfo {title} {Exchange interactions
  among itinerant electrons},\ }in\ \href@noop {} {\emph {\bibinfo {booktitle}
  {Magnetism}}},\ Vol.~\bibinfo {volume} {IV},\ \bibinfo {editor} {edited by\
  \bibinfo {editor} {\bibfnamefont {G.~T.}\ \bibnamefont {Rado}}\ and\ \bibinfo
  {editor} {\bibfnamefont {H.}~\bibnamefont {Suhl}}}\ (\bibinfo  {publisher}
  {Academic Press},\ \bibinfo {address} {New York},\ \bibinfo {year}
  {1966})\BibitemShut {NoStop}%
\bibitem [{\citenamefont {Sandvik}(2010)}]{Sandvik_2010}%
  \BibitemOpen
  \bibfield  {author} {\bibinfo {author} {\bibfnamefont {A.~W.}\ \bibnamefont
  {Sandvik}},\ }\bibfield  {title} {\bibinfo {title} {Computational studies of
  quantum spin systems},\ }in\ \href {https://doi.org/10.1063/1.3518900} {\emph
  {\bibinfo {booktitle} {AIP Conference Proceedings}}},\ Vol.\ \bibinfo
  {volume} {1297}\ (\bibinfo  {publisher} {AIP},\ \bibinfo {year} {2010})\ pp.\
  \bibinfo {pages} {135--338}\BibitemShut {NoStop}%
\bibitem [{\citenamefont {Sylju\aa{}sen}\ and\ \citenamefont
  {Sandvik}(2002)}]{PhysRevE.66.046701}%
  \BibitemOpen
  \bibfield  {author} {\bibinfo {author} {\bibfnamefont {O.~F.}\ \bibnamefont
  {Sylju\aa{}sen}}\ and\ \bibinfo {author} {\bibfnamefont {A.~W.}\ \bibnamefont
  {Sandvik}},\ }\bibfield  {title} {\bibinfo {title} {Quantum {Monte Carlo}
  with directed loops},\ }\href {https://doi.org/10.1103/PhysRevE.66.046701}
  {\bibfield  {journal} {\bibinfo  {journal} {Phys. Rev. E}\ }\textbf {\bibinfo
  {volume} {66}},\ \bibinfo {pages} {046701} (\bibinfo {year}
  {2002})}\BibitemShut {NoStop}%
\bibitem [{\citenamefont {Gmitra}\ and\ \citenamefont
  {Fabian}(2015)}]{PhysRevB.92.155403}%
  \BibitemOpen
  \bibfield  {author} {\bibinfo {author} {\bibfnamefont {M.}~\bibnamefont
  {Gmitra}}\ and\ \bibinfo {author} {\bibfnamefont {J.}~\bibnamefont
  {Fabian}},\ }\bibfield  {title} {\bibinfo {title} {Graphene on
  transition-metal dichalcogenides: A platform for proximity spin-orbit physics
  and optospintronics},\ }\href {https://doi.org/10.1103/PhysRevB.92.155403}
  {\bibfield  {journal} {\bibinfo  {journal} {Phys. Rev. B}\ }\textbf {\bibinfo
  {volume} {92}},\ \bibinfo {pages} {155403} (\bibinfo {year}
  {2015})}\BibitemShut {NoStop}%
\bibitem [{\citenamefont {Wakamura}\ \emph {et~al.}(2019)\citenamefont
  {Wakamura}, \citenamefont {Reale}, \citenamefont {Palczynski}, \citenamefont
  {Zhao}, \citenamefont {Johnson}, \citenamefont {Gu\'eron}, \citenamefont
  {Mattevi}, \citenamefont {Ouerghi},\ and\ \citenamefont
  {Bouchiat}}]{PhysRevB.99.245402}%
  \BibitemOpen
  \bibfield  {author} {\bibinfo {author} {\bibfnamefont {T.}~\bibnamefont
  {Wakamura}}, \bibinfo {author} {\bibfnamefont {F.}~\bibnamefont {Reale}},
  \bibinfo {author} {\bibfnamefont {P.}~\bibnamefont {Palczynski}}, \bibinfo
  {author} {\bibfnamefont {M.~Q.}\ \bibnamefont {Zhao}}, \bibinfo {author}
  {\bibfnamefont {A.~T.~C.}\ \bibnamefont {Johnson}}, \bibinfo {author}
  {\bibfnamefont {S.}~\bibnamefont {Gu\'eron}}, \bibinfo {author}
  {\bibfnamefont {C.}~\bibnamefont {Mattevi}}, \bibinfo {author} {\bibfnamefont
  {A.}~\bibnamefont {Ouerghi}},\ and\ \bibinfo {author} {\bibfnamefont
  {H.}~\bibnamefont {Bouchiat}},\ }\bibfield  {title} {\bibinfo {title}
  {Spin-orbit interaction induced in graphene by transition metal
  dichalcogenides},\ }\href {https://doi.org/10.1103/PhysRevB.99.245402}
  {\bibfield  {journal} {\bibinfo  {journal} {Phys. Rev. B}\ }\textbf {\bibinfo
  {volume} {99}},\ \bibinfo {pages} {245402} (\bibinfo {year}
  {2019})}\BibitemShut {NoStop}%
\bibitem [{\citenamefont {Nelson}\ and\ \citenamefont
  {Kosterlitz}(1977)}]{nelson1977universal}%
  \BibitemOpen
  \bibfield  {author} {\bibinfo {author} {\bibfnamefont {D.~R.}\ \bibnamefont
  {Nelson}}\ and\ \bibinfo {author} {\bibfnamefont {J.~M.}\ \bibnamefont
  {Kosterlitz}},\ }\bibfield  {title} {\bibinfo {title} {Universal jump in the
  superfluid density of two-dimensional superfluids},\ }\href@noop {}
  {\bibfield  {journal} {\bibinfo  {journal} {Phys. Rev. Lett.}\ }\textbf
  {\bibinfo {volume} {39}},\ \bibinfo {pages} {1201} (\bibinfo {year}
  {1977})}\BibitemShut {NoStop}%
\bibitem [{\citenamefont {Flyvbjerg}\ and\ \citenamefont
  {Petersen}(1989)}]{Flyvbjerg1989}%
  \BibitemOpen
  \bibfield  {author} {\bibinfo {author} {\bibfnamefont {H.}~\bibnamefont
  {Flyvbjerg}}\ and\ \bibinfo {author} {\bibfnamefont {H.~G.}\ \bibnamefont
  {Petersen}},\ }\bibfield  {title} {\bibinfo {title} {Error estimates on
  averages of correlated data},\ }\href {https://doi.org/10.1063/1.457480}
  {\bibfield  {journal} {\bibinfo  {journal} {J. Chem. Phys.}\ }\textbf
  {\bibinfo {volume} {91}},\ \bibinfo {pages} {461} (\bibinfo {year}
  {1989})}\BibitemShut {NoStop}%
\bibitem [{\citenamefont {Das}\ and\ \citenamefont
  {Huang}(2024)}]{PhysRevB.109.L060409}%
  \BibitemOpen
  \bibfield  {author} {\bibinfo {author} {\bibfnamefont {M.}~\bibnamefont
  {Das}}\ and\ \bibinfo {author} {\bibfnamefont {C.}~\bibnamefont {Huang}},\
  }\bibfield  {title} {\bibinfo {title} {Unconventional metallic
  ferromagnetism: Nonanalyticity and sign-changing behavior of orbital
  magnetization in rhombohedral trilayer graphene},\ }\href
  {https://doi.org/10.1103/PhysRevB.109.L060409} {\bibfield  {journal}
  {\bibinfo  {journal} {Phys. Rev. B}\ }\textbf {\bibinfo {volume} {109}},\
  \bibinfo {pages} {L060409} (\bibinfo {year} {2024})}\BibitemShut {NoStop}%
\bibitem [{\citenamefont {Wei}\ \emph {et~al.}(2025)\citenamefont {Wei},
  \citenamefont {Xu}, \citenamefont {Villadiego},\ and\ \citenamefont
  {Huang}}]{PhysRevLett.134.046501}%
  \BibitemOpen
  \bibfield  {author} {\bibinfo {author} {\bibfnamefont {N.}~\bibnamefont
  {Wei}}, \bibinfo {author} {\bibfnamefont {G.}~\bibnamefont {Xu}}, \bibinfo
  {author} {\bibfnamefont {I.~S.}\ \bibnamefont {Villadiego}},\ and\ \bibinfo
  {author} {\bibfnamefont {C.}~\bibnamefont {Huang}},\ }\bibfield  {title}
  {\bibinfo {title} {Landau-level mixing and \text{SU}(4) symmetry breaking in
  graphene},\ }\href {https://doi.org/10.1103/PhysRevLett.134.046501}
  {\bibfield  {journal} {\bibinfo  {journal} {Phys. Rev. Lett.}\ }\textbf
  {\bibinfo {volume} {134}},\ \bibinfo {pages} {046501} (\bibinfo {year}
  {2025})}\BibitemShut {NoStop}%
\bibitem [{\citenamefont {Das}\ and\ \citenamefont
  {Huang}(2026)}]{das2025momentum}%
  \BibitemOpen
  \bibfield  {author} {\bibinfo {author} {\bibfnamefont {M.}~\bibnamefont
  {Das}}\ and\ \bibinfo {author} {\bibfnamefont {C.}~\bibnamefont {Huang}},\
  }\bibfield  {title} {\bibinfo {title} {Momentum space ac josephson effect and
  intervalley coherence in multilayer graphene},\ }\href
  {https://doi.org/10.1038/s41467-025-67838-7} {\bibfield  {journal} {\bibinfo
  {journal} {Nat. Commun.}\ }\textbf {\bibinfo {volume} {17}},\ \bibinfo
  {pages} {1079} (\bibinfo {year} {2026})}\BibitemShut {NoStop}%
\bibitem [{\citenamefont {Hsieh}\ \emph {et~al.}(2013)\citenamefont {Hsieh},
  \citenamefont {Kao},\ and\ \citenamefont {Sandvik}}]{hsieh2013finite}%
  \BibitemOpen
  \bibfield  {author} {\bibinfo {author} {\bibfnamefont {Y.-D.}\ \bibnamefont
  {Hsieh}}, \bibinfo {author} {\bibfnamefont {Y.-J.}\ \bibnamefont {Kao}},\
  and\ \bibinfo {author} {\bibfnamefont {A.~W.}\ \bibnamefont {Sandvik}},\
  }\bibfield  {title} {\bibinfo {title} {Finite-size scaling method for the
  berezinskii--kosterlitz--thouless transition},\ }\href@noop {} {\bibfield
  {journal} {\bibinfo  {journal} {J. Stat. Mech.}\ }\textbf {\bibinfo {volume}
  {2013}},\ \bibinfo {pages} {P09001} (\bibinfo {year} {2013})}\BibitemShut
  {NoStop}%
\end{thebibliography}
\end{document}